\documentclass[twocolumn]{aastex631}

\newcommand{\Teff}{$\text{T}_{\text{eff}}$ \ }
\usepackage{amsmath}
\usepackage{apjfonts} 
\usepackage{amsmath,amstext}
\usepackage[T1]{fontenc}
\usepackage{apjfonts} 
\usepackage[figure,figure*]{hypcap}
\usepackage{tablefootnote}
\usepackage{tabularx}
\usepackage{longtable}
\usepackage{threeparttable}

\begin{document}

\title{Modeling Rotation in the Old, Cold Domain: Implications on Gyrochronology and the Stellar Magnetic Wind}

\correspondingauthor{Amanda L. Ash}
\email{ash.172@osu.edu}

\newcommand{\osu}{Department of Astronomy, The Ohio State University, Columbus, 140 W 18th Ave, OH 43210, USA}
\newcommand{\ccapp}{Center for Cosmology and Astroparticle Physics (CCAPP), The Ohio State University, 191 W. Woodruff Ave., Columbus, OH 43210, USA}

\author[0000-0002-8674-9922]{Amanda L. Ash}
\affiliation{\osu}

\author[0000-0003-4769-3273]{Yuxi(Lucy) Lu}
\affiliation{\osu}
\affiliation{\ccapp}

\author[0000-0002-7549-7766]{Marc H. Pinsonneault}
\affiliation{\osu}
\affiliation{\ccapp}




\begin{abstract}

Gyrochronology ties stellar rotation periods to ages. It is well studied in the young, open cluster age domain, but there are few constraints on gyrochronology in the old field star regime. In this work we use gyro-kinematic ages to explore the spin down of stars in this formerly inaccessible domain. Using forward modeling techniques which relax strict Rossby scaled assumptions, we find evidence for a departure from a standard spin down models. This departure can be explained with a mass-dependent term either in the global strength of the stellar wind or in the relationship between angular velocity and wind strength. Models with this additional mass-dependence help explain prior difficulty in fitting open cluster rotation distributions across the full mass range. Additionally, we use rotation models to identify an mass-dependent age domain over which wind driven stellar spin down can be isolated from other physical effects. For lower mass stars, this age domain is most affected by the core-envelope coupling timescale, where as higher mass stars are more subject to inertial effects late in their main sequence lifetimes. Using simple, analytic models of stellar spin down in this domain is unable to determine whether the stellar wind has a mass-dependence or if the stellar wind is strictly Skumanich-like in nature. In future studies of rotational and dynamo evolution we advocate for a forward modeling approach to gyrochronology over purely empirical approaches due to its superior ability to trace physical effects governing stellar spin down.

\end{abstract}

\keywords{Stellar Rotation, Stellar Evolution, Stellar Magnetism, Field Stars}

\section{Introduction}

Stellar ages are crucial for interpreting astrophysical phenomena. At young age scales, the timescales and physics of exoplanet formation and protoplanetary disk dissipation are debated because of disputes surrounding the ages of pre-main sequence stars. At intermediate and late ages, galactic chemical evolution requires accurate stellar chronometers. There are a number of age dating techniques that are suited to different evolutionary phases. Our focus is on gyrochronology, which maps observed rotation periods to stellar ages \citep{2003ApJ...586..464B, 2007ApJ...669.1167B}. This technique is based on observations of stellar spin down, where older star's are observed to rotate slower than their younger counterparts. This technique is especially powerful for low mass stars on the unevolved lower main sequence (MS), as other stellar properties used to infer age change very little over the main sequence lifetime.

Main sequence stars spin down under the influence of a magnetic wind. Using data in the Pleiades, Hyades, and the Sun, , \citet{1972ApJ...171..565S} found a relationship between magnetic activity indicators and stellar angular velocity. This relationship unlocked the possibility of using stellar rotation as an age metric. Additionally, this result was consistent with predictions from the \citet{1967ApJ...148..217W} solar dynamo model, providing insights into the dynamo behavior of younger stars. In this model a star's angular velocity, controls the strength of the magnetic wind which in turn controls the rate of stellar spin down. 

There are domains however where this simple picture of stellar spin down does not hold. At young ages, stars are born with a range of initial conditions, and they maintain the imprint of these initial conditions until they have converged onto a slow rotation sequence. Additionally, stars at young ages are rapidly rotating and are observed to be in the saturated domain \citep{1984ApJ...279..763N}. In this domain, the strict correlation between increasing angular velocity and field strength does not hold, and there is a ''plateau" in the field strength. The magnetically saturated domain is well-studied by young, rapidly rotating stars in open clusters. However, the unsaturated domain isn't as well constrained due to the dearth of old main sequence stars with well-defined ages. 

Stars also need not rotate as solid bodies. At intermediate ages, if the timescale of internal angular momentum transport is longer than the timescale for angular momentum loss, the surface convection zone will spin down faster than the core can respond. When the shears become large enough, angular momentum is from the interior can match the loss from the wind. This proceeds until the core if fully coupled to the envelope. This phenomena therefore includes a transient stalling event in the observed rotation period. This is sometimes referred to as core-envelope de-coupling  \citep{1991ApJ...376..204M, 2010ApJ...716.1269D}. This stage is brief for solar-mass stars, but less massive-partially convective stars can have stalling timescales that extend to several Gyrs \citep{1997ApJ...480..303K, 2020ApJ...904..140C}. Much later in their lifetimes, solar analogs are observed to reach an activity threshold below which spin down becomes inefficient, causing them to rotate faster than standard spin down expectations \citep{2016Natur.529..181V}. 

Finally, wind driven mechanisms and internal angular momentum redistribution are not the sole agents for driving rotational evolution. Stars change in size as they age. As a consequence of angular momentum conservation, stars will spin up or spin down as their inertia changes. As a consequence, there is no apriori reason to expect stars to follow simple, analytic spin down laws. There are domains however where different contributing physical effects will have a minimal impact. It is only in such domains where we can directly map the empirical spin down to a direct measure of the stellar wind properties. 

Open clusters have historically served as the laboratory for calibrating physical and empirical models of stellar spin down as they provide large numbers of rotation periods with robust, rotation-independent ages \citep{2020ApJ...904..140C, 2022ApJ...938..118D, 2013A&A...556A..36G, 2023A&A...672A.159G}. However, most open clusters data is for near solar metallicity stars. Older clusters are also typically more distant, making it difficult to observe lower mass objects. With large time-domain surveys, we have access to large catalogs of older, field star rotation periods \citep{2010Sci...327..977B, Irwin2011, 2019PASP..131a8003M}. However, these datasets generally lack reliable, independent ages. Asteroseismology has been used to get a handful of age-rotation period measurements for stars older than the sun \citep{2015MNRAS.450.1787A, 2024RNAAS...8..260M, 2025ApJ...984..125L}. However, these measurements are sparse, do not reach the lower main sequence, and they provide little constraining power on the late stage evolution of the magnetic wind \cite{2023A&A...672A.159G, 2025ApJ...986...59V}. 

Recent work on gyro-kinematic ages provides an exciting opportunity to establish statistical relationships between rotation period and age for field stars. Gyro-kinematic ages rely on the assumption that stars in the galactic disk experience dynamic heating. If we know a given set of stars share a common age, then their average kinematic properties can be linked to the age of such stars. In older star clusters, the range of rotation periods for a stars of a given mass are observed to converge to a single, slow rotator sequence. By extension, if sufficiently old stars have a common rotation period and HR-diagram position they will be the same age. By taking the average kinematic properties of this subset, one can use the underlying galactic model to define a kinematic age relationship \citep{2020AJ....160...90A, 2021AJ....161..189L} present a catalog of stellar ages inferred using kinematic data. 

The efficacy of these gyro-kinematic ages to trace the dynamic nature of rotational evolution has been previously tested in the literature. \citet{2024AJ....167..159L} fit a fully-empirical gyrochronological model to the data. In doing so, they find their techniques produce comparable age errors to isochronal ages and that they can recover second-order rotation effects such as the  epoch of weakened magnetic braking, consistent with previous studies \citep{2016Natur.529..181V, 2025ApJ...986..120M, 2025ApJ...991L..17M}. There are however limitations in the gyro-kinematic age dating technique. At young ages, the mapping between kinematic properties and age becomes weak as the underlying assumption that stars with similar rotation periods have similar ages breaks down. At older ages, the finite age of the thin disk and systematics between the thin and thick disk limit the viability of gyro-kinematic ages. 

Regardless of the sample population, empirical fits tend to break down when extrapolated beyond the domain in which they are defined, for example to older ages or different metallicities. Additionally, within this empirical framework, it is difficult to disentangle the origins and onset of different stellar physics that affect rotational evolution in a purely empirical approach limiting our ability to understand the underlying dynamo mechanism. Forward models, by contrast can make predictions across a wide range of phase space. However, due to our incomplete knowledge of the physics they require calibration against observations.  

In this study, we expand upon the fully empirical approach by using forward models that consider a set of rotational physics to build gyrochronological relationships for partially-convective stars. We focus on these stars as these stars will converge to the slow rotator sequence at young ages compared to less massive stars. This allows us to avoid having to consider the initial rotation distribution in our models. Our modeling technique allows us to isolate and measure the stellar wind spin down index, the strength of the wind for stars of different mass, and the timescales of angular momentum transport. In particular, we are interested in testing the mass trends predicted by solar-scaled models are correct. Additionally, our techniques allow us to constrain and measure the domain over-which empirical stellar spin down relationships are most well suited. 

In Section \ref{sec: Data} we present the gyro-kinematic datasets and sample selection criteria. The input evolution tracks to the angular momentum evolution models are also discussed in Section \ref{sec: Data}. In Section \ref{sec: models} we present the analytic rotation models and the forward models we fit to the gyro-kinematic sample. The methods for fitting these models is discussed in Sec. \ref{sec:fitting}. In Sections \ref{sec: model results} and \ref{sec: discussion} we present the quantitative results of our model analysis and discuss their implications on the wind properties needed by rotational evolution models. Finally, in Section \ref{sec: conclusion}, we use our results to argue that gyrochronology is best approached using forward modeling techniques as opposed to simple analytic relations.

\section{Data}\label{sec: Data}

In this section we summarize how we combine open cluster and field star rotation periods and ages to establish our observational dataset. This sample spans a range of ages between 10 Myr for the youngest open cluster in the sample and $\approx$ 9 Gyr for the oldest kinematic ages. We sample masses from $0.5 M_{\odot}$ to $1.05 M_{\odot}$ near solar-metallicity, with Gaia XGBoost metallicities between $-0.75 \leq [M/H] \leq 0.75$ \citep{2023ApJS..267....8A}. 

We generate suites of theoretical models using the YREC stellar evolution code \citep{2026arXiv260325792P}. The physical assumptions underlying these models are given in Sec. \ref{sec:input evol}. The predicted rotation periods are generated using the forward modelling code, \verb|Rotevol| with the evolutionary tracks. The data is binned into $0.1 \ M_{\odot}$ bins for the analytic model fittings and $0.05 M_{\odot}$ for the full forward model fits. We elect to use a narrower bin sizes for the full forward models fits as this technique is better able to account for mass-dependent differences in rotational evolution. 

\subsection{Open Cluster Data}

The gyro-kinematic age-rotation catalog is complemented by open cluster period measurements. We use the open cluster rotation catalog presented in \citet{2025ApJ...986...59V} which used rotation measurements in open clusters and stellar associations as a training set for the gyrochronology model \textit{ChronoFlow}. This catalog compiles literature rotation period measurements for 30 open clusters. 

Observations from star forming regions define the initial conditions for our rotation models. At the youngest ages, stars will the circumstellar disk via accretion. By an age of 10 Myr, the large majority of accretion disks have dissipated, making this a logical starting point for our studies \citep{2010A&A...510A..72F, 2013A&A...556A..36G, 2016ApJ...829...32S}. Our models are initialized at $10$ Myr with an initial rotation period of $8$ days, based on observations of Upper Scorpius, which has a period distribution between $\approx 2 - 8$ days for partially-convective stars \citep{2017ApJ...850..134S}.  

The older clusters in this sample provide good overlap with the gyro-kinematic age sample near $1$ Gyr. The younger clusters capture rotational evolution at ages than the gyro-kinematic ages are not well-suited for. This is a vital domain to provide constraints for our model fits because behavior in the young cluster domain can affect the trajectory of rotational evolution at older ages. 

\subsection{Field Star Data}

We compliment the open cluster sample with gyro-kinematic ages for field stars. Gyro-kinematic ages are calculated following the method described in \citet{2021AJ....161..189L}, by measuring the vertical velocity dispersion, $\sigma_{vz}$, for stars of similar rotation period measurements, photometric temperature, and absolute Gaia $G$-band magnitude.
The rotation periods include measurements from Kepler \citep{2010Sci...327..977B, McQuillan2014, Santos2019}, MEarth \citep{Irwin2011, Newton2017}, and ZTF \citep{2019PASP..131a8003M, 2012AJ....144..145B, 2022AJ....164..251L, 2024AJ....167..159L, 2024NatAs...8..223L}. Binaries are excluded from the sample by excluding stars with RUWE values greater than $1.4$ and equal mass binaries are excluded based on a magnitude cut of 0.6 mag. 

Stars with rotation periods less than $10$ days can be members of synchronized binary systems and will have a different spin down history than single star systems\citep{2017AJ....154..250L}. While single star systems can rotate this rapidly, we exclude stars in the field star sample that rotate more rapidly than this limit as these stars are quite young and well represented by the open cluster sample. Additionally, we do not include measurements from TESS \citep{TESS} as it can only obtain periods $<$ 14 days, which leaves only a narrow domain where we can obtain useful data for our purposes. If TESS measurements are included, there is an apparent discontinuity between the cluster age scale and the gyro-kinematic age scale. We do not recommend future studies using the gyro-kinematic ages use ages measured from TESS data. 

We use extinction dust maps from {\tt dustmap} and Gaia parallaxes to correct for absolute G-band magnitude \citep{Green2018, Green20182}. The photometric surface temperatures are computed using the extinction corrected Gaia Bp-Rp colors and polynomial fits taken from \citet{2020ApJ...904..140C}.

\subsection{Input Evolution Models}\label{sec:input evol}

The mass tracks used to forward model the rotational evolution and build the observed mass sequences are computed with the Yale Rotation Evolution Code (YREC) (\citet{1989ApJ...338..424P, 2026arXiv260325792P}). These tracks use a standard mixing length prescription with a solar calibrated mixing length parameter of $ \alpha_{mlt}  = 1.92$ and a helium mass fraction of $Y = 0.26$ \citep{1968pss..book.....C}. Opacities are computed using OP opacity tables with a solar mixture from \citet{2005MNRAS.360..458B, 1998SSRv...85..161G} and low temperature molecular opacities from \citet{2005ApJ...623..585F}. The equation of state is given by the OPAL EoS \citep{1996ApJ...456..902R, 2002ApJ...576.1064R}, and we use the \citet{1992IAUS..149..225K} model atmospheres to define the surface boundary conditions. The models do not consider diffusion or additional mixing. We apply bolometric corrections to the predicted luminosities from the evolutionary model to determine the theoretical G-band magnitudes using tables from \citet{2023A&A...674A..26C}. These corrections have a magnitude error of $\approx 0.015$ mag.

\section{Rotation Models}\label{sec: models}

In this section, we summarize the physics and assumptions used in our angular momentum evolution models. We also discuss how enforcing strict assumptions allows the physical model to be translated into analytic forms used in empirical gyrochronology. Finally, we use our forward models to define an age domain over which stellar spin down will be wind-dominated.  

\subsection{Forward Models}\label{sec: forward models}

To model the rotational evolution of partially-convective stars, we use the rotational evolution code, \verb|Rotevol| in tandem with the YREC stellar evolution code \citep{1989ApJ...338..424P, 2026arXiv260325792P}. This code takes the structure from a non-rotating YREC model as input and computes the rotational evolution with a given set of rotational physics and initial conditions \citep{1989ApJ...338..424P}. The physics of our input evolution tracks are discussed in Sec. \ref{sec: Data}. 

\verb|Rotevol| does not compute the feedback between stellar structure, rotation and magnetic effects such as starspots. However, as shown in \citet{2026arXiv260325792P}, the predictions from a non-solid body \verb|Rotevol| computation and a self-consistent rotating YREC model are close for older, less active stars. The greatest deviations between the two models are for very active young stars, which are not our central focus. Further, we elect to use \verb|Rotevol| as it is orders of magnitude faster than a full evolutionary model and allows us to investigate a large parameter space. 

In the case where a star launches a magnetic wind and experiences inertial changes, the change in the angular velocity is given by: 

\begin{equation}\label{eq: angular velocity change}
    \frac{\dot{\omega}}{\omega} = \frac{\omega_0}{\omega} \frac{\dot{I_*}}{I_*} + \frac{\zeta \dot{M_*} R_A^2}{I} . 
\end{equation}

where $I$ the moment of inertia, $\dot{I}$ is the rate of change in the inertia, and $\dot{M_*}$ is the mass loss rate of the stellar wind. For solid body rotation, the moment of inertia is $k M_* R_*^2$. $k$ represents the distribution of mass in the star and is typically a value of $0.07$ for solar-like main sequence stars. In the more general case one would need to solve for internal transport of angular momentum and shears induced by structural evolution. The stellar wind torque is caused by co-rotation between a star and material in the wind up to the Alfvén radius, the point where the magnetic energy density is equal to the kinetic energy density. The Alfvén radius depends on the strength of the stellar magnetic field, the mass loss rate, and the wind speed. The $\zeta$ pre-factor in the second term of Eq. \ref{eq: angular velocity change} describes the geometry of the wind. 

Our wind model is adopted from \citet{2012ApJ...754L..26M} which used MHD simulations to parameterize the stellar torque in terms of the mass loss rate, the stellar magnetic field strength, and the angular velocity: 

\begin{equation}
\tau_{wind}  = \frac{K_1^2}{(2G)^m}B^{4m} \frac{dM}{dt}^{1 - 2m} \frac{R^{5m + 2}}{M^m}\frac{\omega}{K_2^2 + \frac{\omega^2 R^3}{2 G M}}
\end{equation}

where $K_1 = 1.30$ and $K_2 = 0.056$. The $m$-parameter, equal to 0.22, is the effective efficiency of the stellar wind determined by fits to the outputs of MHD simulations \citep{2012ApJ...754L..26M}.

The model includes terms for structural evolution and a centrifugal correction term to the $K_2$ constant. The final term in the equation is a centrifugal correction term and becomes important for rapid rotation, which is a short-lived phase for the stars of interest. The structure term is given accounts for inertial changes induced by mass or radius changes.

We use scalings to the solar mass loss rate and the solar field strength to compute the wind torques in our model, as direct measurements are in general not possible. Our scalings take the form described in \citet{2013ApJ...776...67V}. The mass loss rate is given by: 

\begin{equation}\label{eq:mass loss rate}
\frac{\dot{M}}{\dot{M_{\odot}}} = 
    \begin{cases}
    \bigg(\frac{Ro}{Ro_{\odot}}\bigg)^{-a} \frac{L}{L_{\odot}} ; \  Ro  \ <  \ Ro_{crit}\\
    \bigg(\frac{Ro_{crit}}{Ro_{\odot}}\bigg)^{-a} \frac{L}{L_{\odot}} ; \  Ro  \ >  \ Ro_{crit}\\
    \end{cases}
\end{equation}

and the field strength scaling is given by:

\begin{equation}\label{eq:field scaling}
\frac{B}{B_{\odot}} = 
    \begin{cases}
    \bigg(\frac{Ro}{Ro_{\odot}}\bigg)^{-b} \bigg(\frac{P_{phot}}{P_{phot, \odot}}\bigg)^{\frac{1}{2}} ; \  Ro  \ <  \ Ro_{crit}\\
   \bigg(\frac{Ro_{crit}}{Ro_{\odot}}\bigg)^{-b} \bigg(\frac{P_{phot}}{P_{phot, \odot}}\bigg)^{\frac{1}{2}} ; \  Ro  \ >  \ Ro_{crit}\\
    \end{cases} .
\end{equation}

The solar mass loss rate is low, losing mass at a rate of $2  \times 10^{-14} M_{\odot} yr^{-1}$ \citep{2019ApJ...885L..30F}. At this limit, we must rely on other proxies to infer it such as the coronal heating rate. \citet{2005ApJ...628L.143W} demonstrate that there is an empirical relationship between the mass loss rate and the X-ray luminosity of a star. 

The magnetic field term is derived by assuming that $B \propto f_{spot} B_{eq}$, where $B_{eq} \propto P_{phot}^{0.5}$. 

Both of the mass loss rate and field strength scalings are parameterized in terms of the Rossby number:  

\begin{equation} \label{eq:Ro}
    Ro = \frac{P_{rot}}{\tau_{CZ}}, 
\end{equation}

where $P_{rot}$ is the stellar rotation period and $\tau_{CZ}$ is the convective turnover timescale. Empirically, stellar activity in stars across masses is more strongly correlated with Rossby number than rotation period alone \citep{1984ApJ...287..769N}. 

Neither the the mass loss rate nor the field strength can be arbitrarily large, represented by the critical Rossby number in Eqs. \ref{eq:mass loss rate} and \ref{eq:field scaling}. The x-ray luminosity has been empirically demonstrated to scale with the Rossby number up to some saturation threshold \citep{2005ApJ...628L.143W}. Similarly, field strengths and activity proxies are also shown to saturate \citep{2003A&A...397..147P}. Magnetic saturation has been shown to occur near a Rossby number of $\approx 0.13$ \citep{2013AN....334..151W}. Additionally, rapidly rotating stars in young open clusters implies that torques cannot be arbitrarily large. These observation set a critical angular velocity threshold for saturation: 

\begin{equation} \label{eq:saturation}
    \frac{\omega_{crit}}{\omega_{\odot}} = K   \bigg(\frac{\tau_{cz, \odot}}{\tau_{cz}}\bigg) . 
\end{equation}

As the x-ray luminosity is associated with the coronal magnetic field strength, it is assumed that the field strength and the mass loss rate share a common saturation threshold.

The full functional form for the angular momentum evolution is then given by:

\begin{equation}\label{eq: Jdot}
\frac{\dot{J}}{\dot{J_{\odot}}} = 
\begin{cases}
  \begin{aligned}
    F_k \frac{\omega}{K_2^2 + \frac{\omega^2 R^3}{2 G M}} \bigg(\frac{S}{S_{\odot}} \bigg)&\bigg( \frac{\omega}{\omega_{\odot}} \bigg)^{4mb + a -2ma + 1}\\&\bigg( \frac{\tau_{CZ}}{\tau_{CZ, \odot}} \bigg)^{4mb + a -2ma}
  \end{aligned} & ; Ro < Ro_{crit} \\
  \begin{aligned}
  F_k \frac{\omega}{K_2^2 + \frac{\omega^2 R^3}{2 G M}} \bigg(\frac{S}{S_{\odot}} \bigg) &\bigg( \frac{\omega_{crit}}{\omega_{\odot}} \bigg)^{4mb + a -2ma + 1}\\&\bigg( \frac{\tau_{CZ}}{\tau_{CZ, \odot}} \bigg)^{4mb + a -2ma}
  \end{aligned} & ; Ro > Ro_{crit}
\end{cases}
\end{equation}

The structure terms in the angular momentum evolution are represented by $\frac{S}{S_\odot}$ and is given by: 

\begin{equation}\label{eq: structure}
\frac{S}{S_{\odot}} = \bigg( \frac{R}{R_{\odot}} \bigg)^{2m + 5} \bigg( \frac{M}{M_{\odot}} \bigg)^{\neg m} \bigg( \frac{L}{L_{\odot}} \bigg)^{1 - 2m} \bigg( \frac{P_{atm}}{P_{atm, \odot}} \bigg)^{2m} . 
\end{equation}
    
$F_k$ controls the overall strength of the torque and is typically used to calibrate a spin down model to reproduce the solar rotation rate at solar age. $F_c$ encapsulates deviations from spherical symmetry due to rapid rotation. For non-rapid rotators, this is a value nearly equal to $1$ and becomes important for rapid rotation, which is a short-lived phase for the stars of interest. The structure term is given accounts for inertial changes induced by mass or radius changes. The $\alpha_{wind}$ exponent is the spin down rate if there were no inertial effects. Generically, one can relate the instantaneous spin down rate to the angular velocity as $\dot{\omega} \propto {\omega^{1 + \frac{1}{\alpha_{wind}}}}$. This relation allows for the time-dependent wind index to be solved for in terms of the free model parameters: 

\begin{equation}\label{eq:windparameters}
    \alpha_{wind} = \frac{1}{4mb + a - 2ma}. 
\end{equation}

Beyond the wind prescription, the model includes terms for structural evolution and a centrifugal correction term. The latter, 

\begin{figure*}
    \centering
    \includegraphics[width=0.99\textwidth]{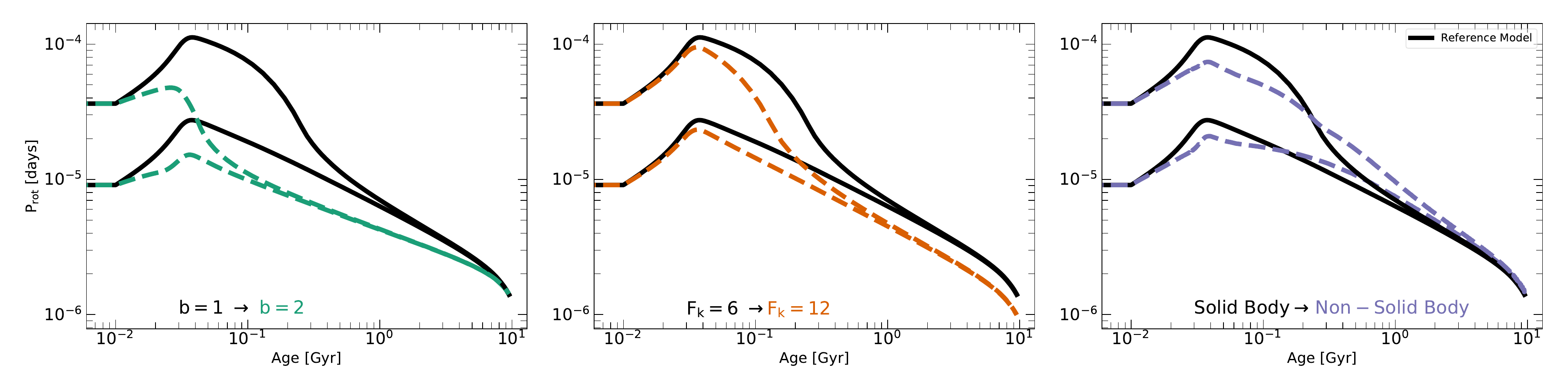}
    \caption{Demonstration of forward model parameters. Model comparisons are made against the ''standard" models given in solid black lines. Models are launched with an initial rotation period of 8 days and 2 days. Left Panel: Changes in the b parameter (dashed), changes the strength of the magnetized wind and mostly affects the slope of the spin down in the unsaturated, or older domain. Middle: Adjusting the torque scaling factor (dashed) corresponds to a translation in the angular velocity, but the slope of the spin down remains the same. Right: Difference between a solid body model (solid) and a non-solid body model with a core-envelope coupling timescale of $\tau_{CE} = 30 \ Myr$.}
    \label{fig:model demo}
\end{figure*}

Thus far, we have assumed that the star is uniformly rotating which is not in general true. Observations of partially-convective stars reveal three distinct spin down phases. In the first phase, the envelope will spin down independently. In the second phase, there will be a mass-dependent ''stalling" where the core and the envelope couple. Once the core has had time to respond, to the spin down of the envelope, the core will act as an angular momentum source for the envelope causing a lapse in the observed spin down of the envelope as it gets additional rotational energy from the core. Finally, after the core and the envelope couple, the star will resume stellar spin down as a solid body. Core-envelope de-coupling is observed in open clusters and follows a mass-dependent timescale. 

As internal angular momentum physics is not the central focus of this work, we adopt a two-zone approximation for core-envelope coupling as opposed to a self-consistent, continuous calculation. This model was first introduced in \citet{1991ApJ...376..204M} and describes the angular momentum of the stellar core and envelope as: 

\begin{equation}
    \dot{J_{core}} = -\frac{\Delta J}{\tau_C} ,
\end{equation}

and 

\begin{equation}
    \dot{J_{envelope}} = \frac{\Delta J}{\tau_C} - \frac{J_{conv}}{\tau_J}.
\end{equation}

$\Delta J$ is the difference between the angular momentum of the core and envelope, $J_{conv}$ is the angular momentum held in the convective envelope, $\tau_C$ is the timescale over which angular momentum will be transferred between the core and the envelope, and $\tau_J$ is the characteristic timescale for angular momentum loss from the wind. 

Figure \ref{fig:model demo} shows the effect of different model parameter changes on the predicted angular velocity of a solar-mass star initialized with a slow rotation period and a rapid one. To provide a comparison point, our models are shown against a standard wind model which uses a solar-scaled torque in the VSP13 framework. This model assumes that $a = 2$, $b = 1$, $m = 0.22$, and $F_k = 6.571$. Changes in the field index, b, of the star induce a change in the slope of spin down. The torque scaling parameter, $F_k$ changes the zero-point of the rotation-age relationship, but not the slope. The core-envelope coupling timescale will act to temper the spin down of the star. At young ages, the spin down rate will be suppressed such that the predicted rotation periods are shorter than a solid body model. At later ages, the star will not have spun down as much as anticipated and a solid body model will have a shorter rotation period than a non-solid body one. At the oldest ages, the slope of a model which includes core-envelope de-coupling will revert to that of a solid-body model.

While the Rossby scalings broadly describe the spin down behavior of stars, they may not be a complete representation of all the physics at play. The convective turnover timescale is not a direct observable and how one defines the convective turnover time imposes a functional form on the torque that is model-dependent. More broadly, the wind model relies on a series of assumed scalings which should be critically examined - particularly when they are extrapolated beyond the regime in which they are calibrated. We therefore explore how relaxing assumption in the wind torque model affect the interpretation of spin down in the partially-convective domain. 

Using these model physics, we adopt four scenarios with different permutations of free parameters to fit against the mass sequences constructed in Section \ref{sec: Data}. The initial rotation period is set to 8 days with a disk locking lifetime of 10 Myr. 

To reproduce the observed Rossby saturation threshold, we set $K = 10$ in Eq. \ref{eq:saturation}. In the mass range of interest, all stars drop below the saturation threshold prior to an age of $\approx 1 Gyr$. As the stars we are interested in are considerably older, in the age range between $2-10$ Gyr, our dataset has little constraining power on the saturation threshold. 

In our first model scenario, we assume the highest degree of freedom where the stellar wind index, the torque scaling factor, and core-envelope coupling timescale are allowed to vary with mass. In our models, $a$ and $b$ are degenerate. We therefore allow $b$ to be the free parameter which controls the stellar wind index, $\alpha_{wind}$ while $a$ is fixed. The mass loss index, $a$ and the field efficiency parameter, $m$ are always set to values of $2$ and $0.22$, respectively. In the second scenario, we hold $b$ fixed as well and the remaining parameters, $F_k$ and $\tau_{CE}$ remain free. We also fit a solid body model version of this model where $F_k$ is the only free parameter. In the final scenario, we constrain the torque scaling factor to a value which reproduces the solar rotation period at the solar age solid-body rotation model with a Skumanich-like wind index and allow all other parameters to be free. The properties of these model scenarios are given in Table \ref{tab: model params}.

\begin{table*}
\footnotesize
    \centering
    \begin{threeparttable}
    
    \begin{tabular}{c c c c c c c c c c }
         Model Name & a & b & m & $F_k$ & $\tau_{CE}$ & $\omega_{sat}$ & $P_{rot, 0}$ & $\tau_{disk}$ \\
         \hline
         All free & 2 & 0.5-5 & 0.22 & 1-100 & 1-14 Gyr & $2.83\times10^{-5} \ rad \ s^{-1}$  & $8 \ days$ & $10 \ Myr$ \\
         Wind Constrained & 2 & 1 & 0.22 & 1-100 & 1-14 Gyr & $2.83\times10^{-5} \ rad \ s^{-1}$  & $8 \ days$ & $10 \ Myr$ \\
         Torque Constrained & 2 & 0.5-5 & 0.22 & 6.571\tnote{1} & 1-14 Gyr & $2.83\times10^{-5} \ rad \ s^{-1}$  & $8 \ days$ & $10 \ Myr$ \\
         Solid Body & 2 & 1 & 0.22 & 1 - 100 & - & $2.83\times10^{-5}\ rad \ s^{-1}$  & $8 \ days$ & $10 \ Myr$ \\

        \hline
    
    \end{tabular}

    \begin{tablenotes}
    \item[1] Solar-calibrated torque scaling value. This $F_k$ value reproduces the solar angular velocity ($2.83 \times 10^{-6}$ at the solar age (4.568 Gyr).)
        
    \end{tablenotes}
    
    \caption{Model parameter choices}
    \label{tab: model params}

    \end{threeparttable}
\end{table*}

There are advantages to both forward modeling techniques and empirical relationships. We therefore tie our forward models and the assumptions taken there within to commonly used analytic prescriptions in Sec. \ref{sec: analytic models}. 

\subsection{Connection to Empirical Models}

The empirical slope of spin down in the age-rotation relationship depends on many physical processes. To analytically measure the wind spin down index and link the forward models to empirical gyrochronology, we must isolate the wind from other physical effects. 

\begin{figure}\label{fig:IC convergence}
    \centering
    \includegraphics[width=0.5\textwidth]{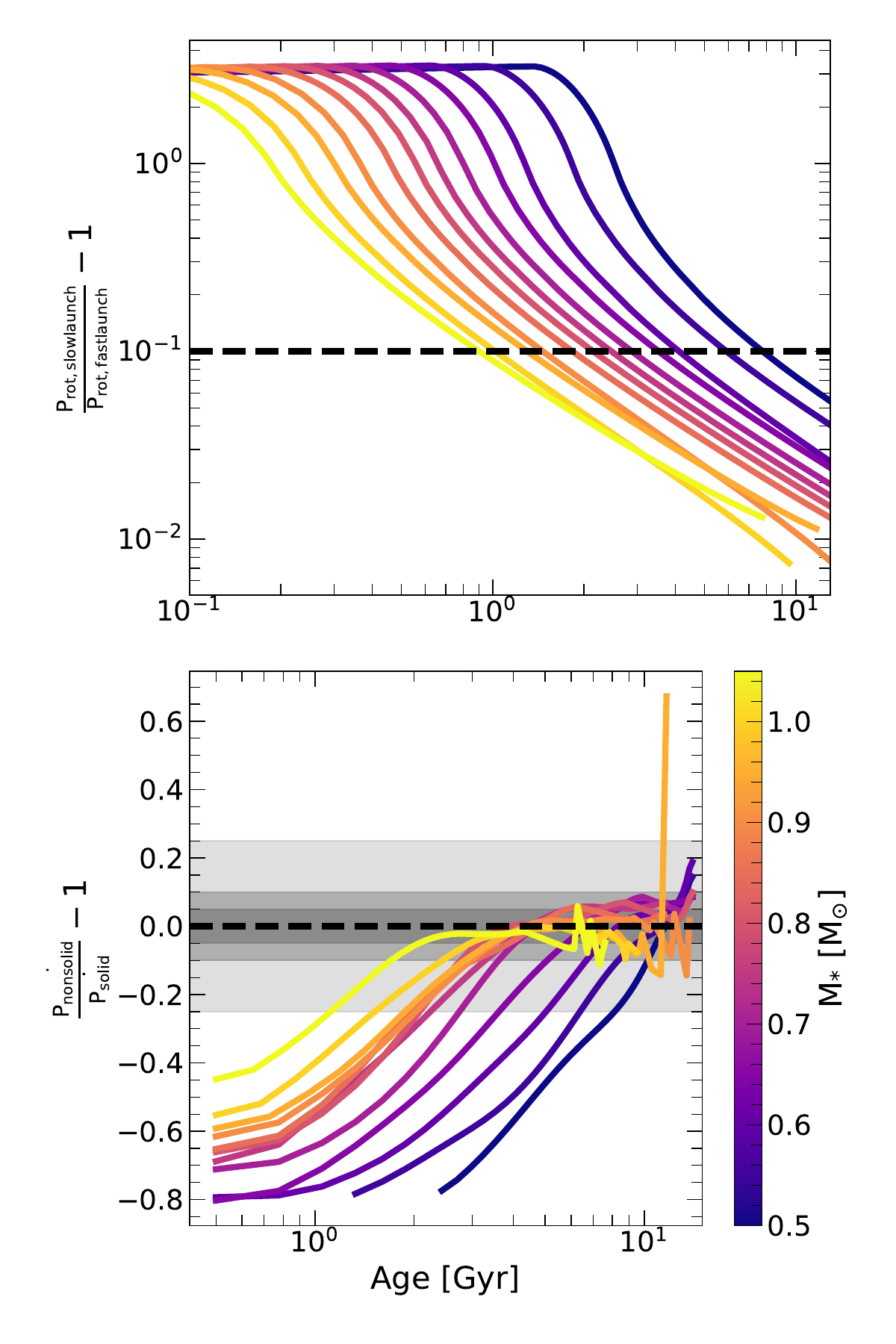}
    \caption{Convergence timescale for different effects. Upper panel: Fractional difference between rotation periods from models launched with an initial period of 2 days and 8 days, which is a representative range for 10 Myr stars. Black dashed lines shows when the model rotation periods agree to within 10 \%. Lower panel: Fractional difference in the slope of spin down between a solid body and non-solid body model. Shaded regions show the convergence in measured spin down at the 20 \%, 10\%, and 5\% level from lightest to darkest, respectively. For this panel, the initial rotation period for all models is $8$ days}
    \label{fig:placeholder}
\end{figure}

An absolute lower age limit where one can begin to tease out stellar wind physics is set by the age at which stars will converge onto a single, slow rotation sequence. Using the initial conditions discussed in Sec. \ref{sec: Data}, we define the converged age to be the age at which the predicted rotation periods from a model launched with an initial period of $2$ days and one at $8$ days agree to within $10 \%$. The upper panel of Fig. \ref{fig:IC convergence} shows how this metric changes with different mass tracks.

\begin{figure*}\label{fig:interior_contribution}
    \centering
    \includegraphics[width=\textwidth]{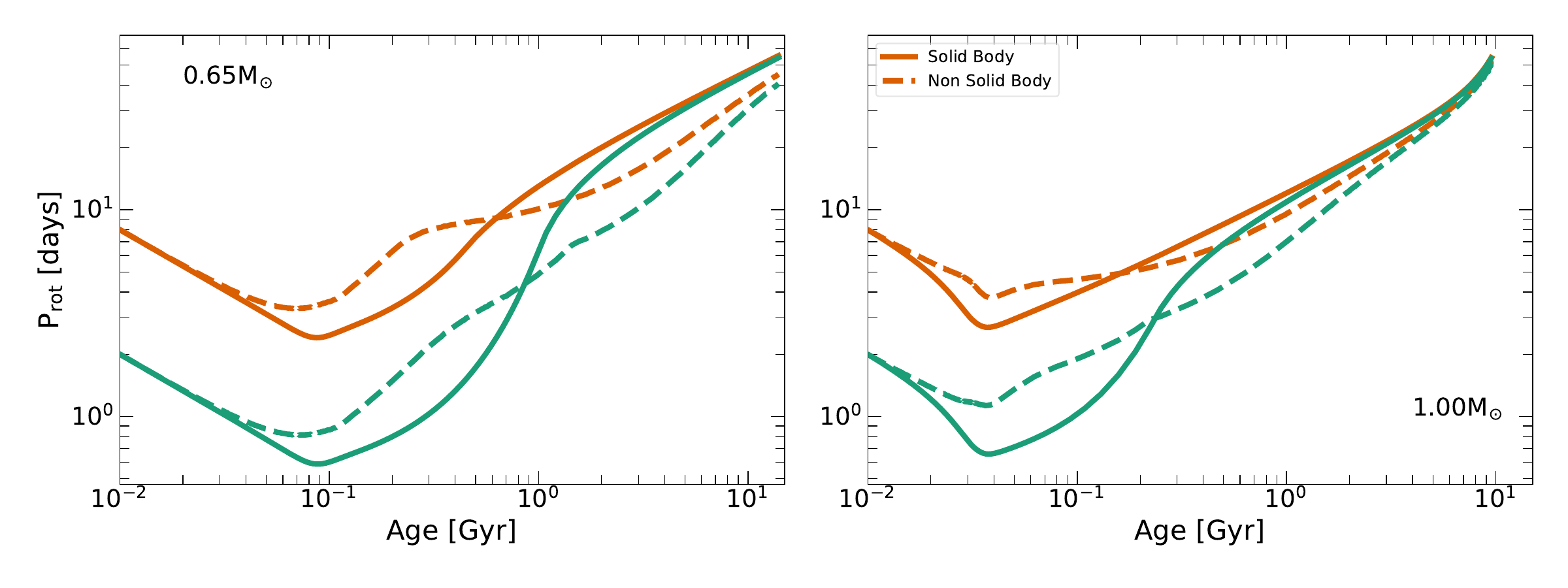}
    \caption{Difference between solid-body and non-solid body evolution for stars with different masses and initial rotation periods. Left: Solid-body and Non-solid body rotation tracks for a lower mass $0.65 M_{\odot}$ star launched with a rotation period of $8$ days (orange) and $2$ days (teal). The core-envelope coupling timescale is set to a mass-dependent value of 388 Myr. Right: Same as left panel, but using a solar-mass track. The coupling timescale is set to $30$ Myr. }
    \label{fig:placeholder}
\end{figure*}

For non-solid body models, the lower age limit for the wind dominated domain is set by the core-envelope coupling timescale. We compute solid and non-solid body rotational evolution tracks for stars in our mass range of interest. Using a standard wind prescription and core-envelope coupling timescales given by \citet{2016ApJ...829...32S}, we compute the difference in the first-derivative between a solid-body model and a non-solid body model. We use this metric as it is most closely ties to the stellar spin down index. Figure \ref{fig:interior_contribution} shows a comparison between solid-body and non-solid body models for different masses and Fig. \ref{fig:IC convergence} shows the fractional difference between the time derivative of rotation as a function of age. We define the lower limit of the wind-dominated domain to be the age at which the absolute value of the fractional difference between the solid body and non-solid body model is less than $10\%$ This lower age limit has a significant impact on the interpretation of empirical spin down measurements of low mass stars and is discussed in further detail in Sec. \ref{sec: wind dominated domain}.  

\begin{figure*}\label{fig: intertialspindown}
    \centering
    \includegraphics[width=\textwidth]{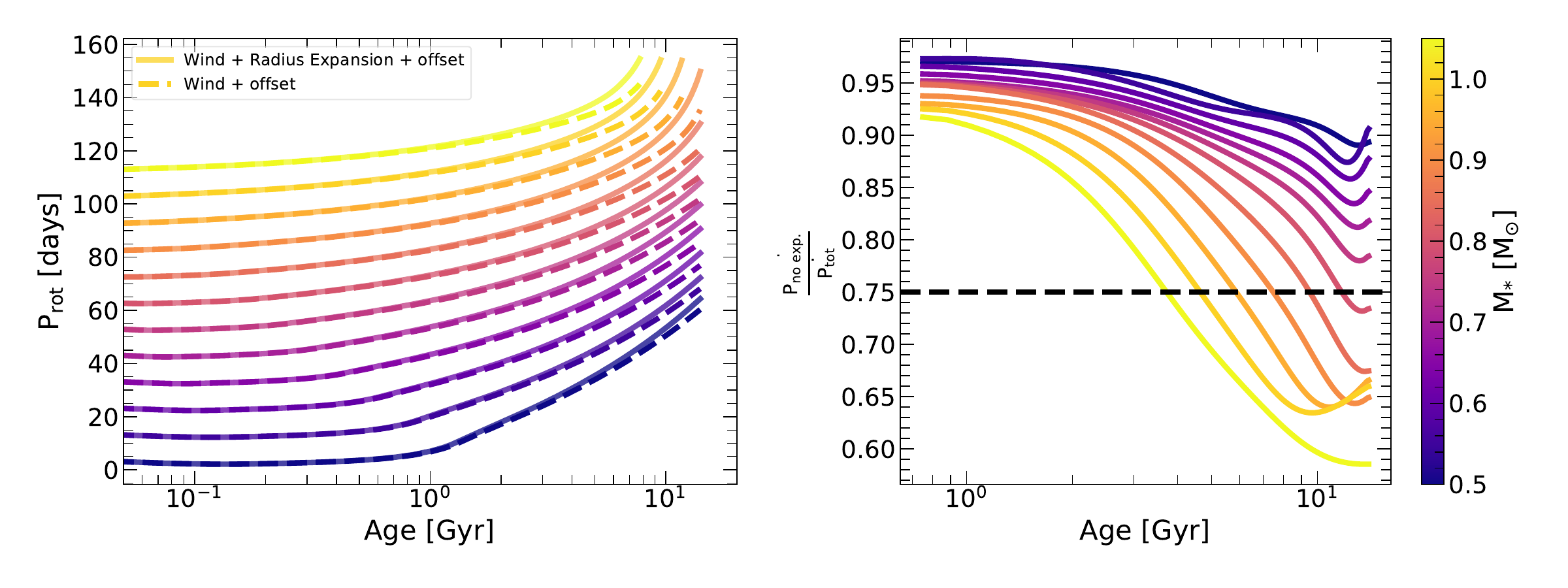}
    \caption{Inertial Contributions to stellar spin down. Left: Predicted rotation periods for a rotation models which computes stellar wind torques and ones that only consider structural changes of the star. Right: Ratio between the time derivative of the rotation period for models given in the left panel. The dashed black lines show when the inertial component contributes $1\%$, $5\%$ and $10\%$ of the total spin down. Where the solid lines cross these thresholds determines the age where radius expansion is expected to contribute some percentage to the total spin down. }
    \label{fig:placeholder}
\end{figure*}

\begin{figure}
    \centering
    \includegraphics[width=0.5\textwidth]{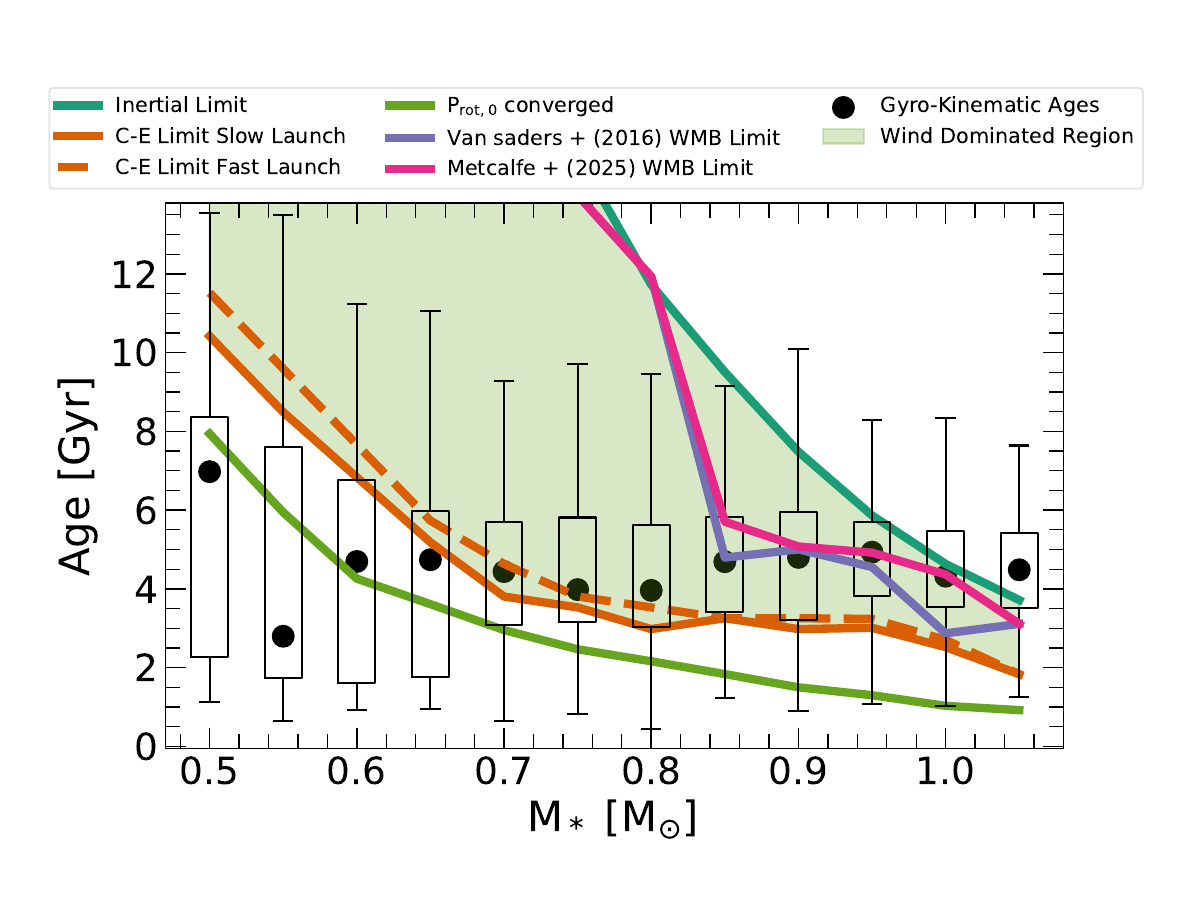}
    \caption{Wind-dominated ages for different stellar masses. The shaded green region shows where one would expect a star's spin down to be dominated by stellar wind torques. The lower and upper age limits for this region are determined by the thresholds set by Figures \ref{fig:interior_contribution} and  \ref{fig: intertialspindown}, which determine the differences between the predicted spin down of models with varying assumptions. Also featured on this figure is the age at which we expect the weakened magnetic braking epoch to occur as a function of mass. Finally, the black points denote the median age of a given mass sample with the box plot showing the rest of the age distribution.}
    \label{fig: wind dominated region}
\end{figure}

The upper age limit is set by structural evolution. Higher mass stars will experience a greater degree of radius expansion on the main sequence. From the rotation tracks, we measure the time derivative of the rotation period predicted from the full model and compare it to the limiting case where there is no inertial changes and only the stellar wind torque contributes to spin down. Figure \ref{fig: intertialspindown} shows the ratio between these two cases for stars of different masses. This figure indicates that radius expansion will have a larger impact earlier in the main sequence lifetime for higher mass stars. We summarize these age bounds in  Fig. \ref{fig: wind dominated region}. 

We highlight that the stars between $0.70 M_{\odot}$ and $0.90 M_{\odot}$ have the longest wind-dominated phase. These masses represent a ''sweet spot" for our model fittings and we pay special attention to this domain in later sections.

\subsection{Analytic Models}\label{sec: analytic models}

A physical forward model allows us to disentangle the physics governing stellar spin down. However, analytic relationships are commonly used in the literature and serve as a basis of gyrochronology. To determine whether commonly used analytic models can be extrapolated from the young cluster domain to the older field stars, we fit three simple analytic models to the mass sequences described in Sec. \ref{sec: Data}. The first two are a classical spin down approach with a free spin down index and zero-point: 

\begin{equation}\label{eq: const spin}
    P_{rot}(t) = P_{rot, \ 1Gyr} \bigg(\frac{t}{1 \ Gyr}\bigg)^{\alpha} .
\end{equation}

The term, $P_{rot, \ 1Gyr}$, anchors the predicted rotation periods to the rotation period at $1$ Gyr. In the first model case, we fix the spin down index to be Skumanich, such that $\alpha = 0.5$. In the second model, we allow both the spin down index and the zero-point to be free. 

Finally, we consider a broken-power law model which is physically motivated by work done by \citet{2016Natur.529..181V} on the phenomena of weakened magnetic braking. Our analytic model for this scenario is given by:

\begin{equation}\label{eq:weakened magnetic braking}
    P_{rot}(t) = 
\left\{
    \begin{array}{lr}
        P_{rot, \ 1Gyr} \bigg(\frac{t}{1 \ Gyr}\bigg)^{\alpha_1}, &  t \le t_{break}\\
        P_{rot, \ t_{break}} \bigg(\frac{t}{t_{break}}\bigg)^{\alpha_2}, &  t \le t_{break}
    \end{array}
\right\}
\end{equation}

where $P_{rot, \ 1Gyr}$, $t_{break}$, $\alpha_1$, and $\alpha_2$ are free parameters. $t_{break}$ is the age at which the star shifts to a new spin down index. $\alpha_1$ and $\alpha_2$ are the spin down indices before and after the star reaches an age of $t_{break}$, respectively. We also assign a prior where $\alpha_2$ must be less than or equal to $\alpha_1$. 

These analytic models are fit to the wind dominated domain defined in Section \ref{sec: wind dominated domain}, such that the best fit spin down indices of these models should reflect wind driven spin down.

\section{Model Fittings}\label{sec:fitting}

In this section we summarize the techniques used to construct and fit rotation sequences as a function of mass. 

\subsection{Mass Sequences}

From the gyro-kinematic and open cluster data, we construct evolutionary sequences for stars of a given mass. By using mass as the independent variable, we are able to relax the Rossby assumption which will impose a functional form on the mass-dependence of stellar spin down. This allows us to solve for the ''true" mass-dependence in different spin down properties. Additionally, we are better able to mitigate biases in our model fittings induced by stars which share observational properties, but are in different evolutionary states. This is in contrast to many other literature studies which commonly use \Teff or color the main stellar coordinate to investigate stellar spin down (e.g. \citet{2024AJ....167..159L, 2016ApJ...823...16B, 2022ApJ...938..118D}). 

To select for mass sequences, we find stars which have \Teff values and Gaia G-band magnitudes consistent with evolutionary model predictions described in Sec. \ref{sec:input evol}. We compute magnitude errors on the observed G-band magnitudes using the Gaia parallax errors and G-band flux errors. These errors age given by: 

\begin{equation}
    \sigma_{G} = \sqrt{\sigma_m^2 + (\frac{5}{p} \sigma_p)^2}, 
\end{equation}

where $p$ and $\sigma_p$ represent the parallax and the error on the parallax. $\sigma_m$ is the error on the apparent magnitude and is approximated by: 

\begin{equation}
    \sigma_{m} \approx 1.0857 \frac{\sigma f_G}{f_G} 
\end{equation}

where $f_g$ is the flux measured in the Gaia G-band. The average magnitude error in the sample is $\approx 0.05$ mag. 

We select stars for our mono-mass sequences with a generous $5 \sigma$ magnitude cut a for a star of a given \Teff . We select stars for a given mass sequence by requiring a star's observed G-band magnitude to be within $5 \sigma$ of the predicted magnitude, using an interpolation table of \Teff -G relations predicted by the models. We select for mass sequences between $0.5 \ M_{\odot} - 1.05 \  M_{\odot}$ in steps of $0.05 \ M_{\odot}$. This selection technique is applied to both the gyro-kinematic dataset and the open cluster catalog. 

In addition to making cuts based on solar metallicity mass tracks, we further isolate stars using $3 \sigma$ outlier cut on $[M/H]$ as measured from \citet{2023ApJS..267....8A}. After selection, our field sample has a mean metallicity of $-0.08 \pm 0.15$. Finally, using a rolling median, we exclude stars that are $3 \sigma$ outliers in $P_{rot}$.  

\begin{figure*}
    \centering
    \includegraphics[width=\textwidth]{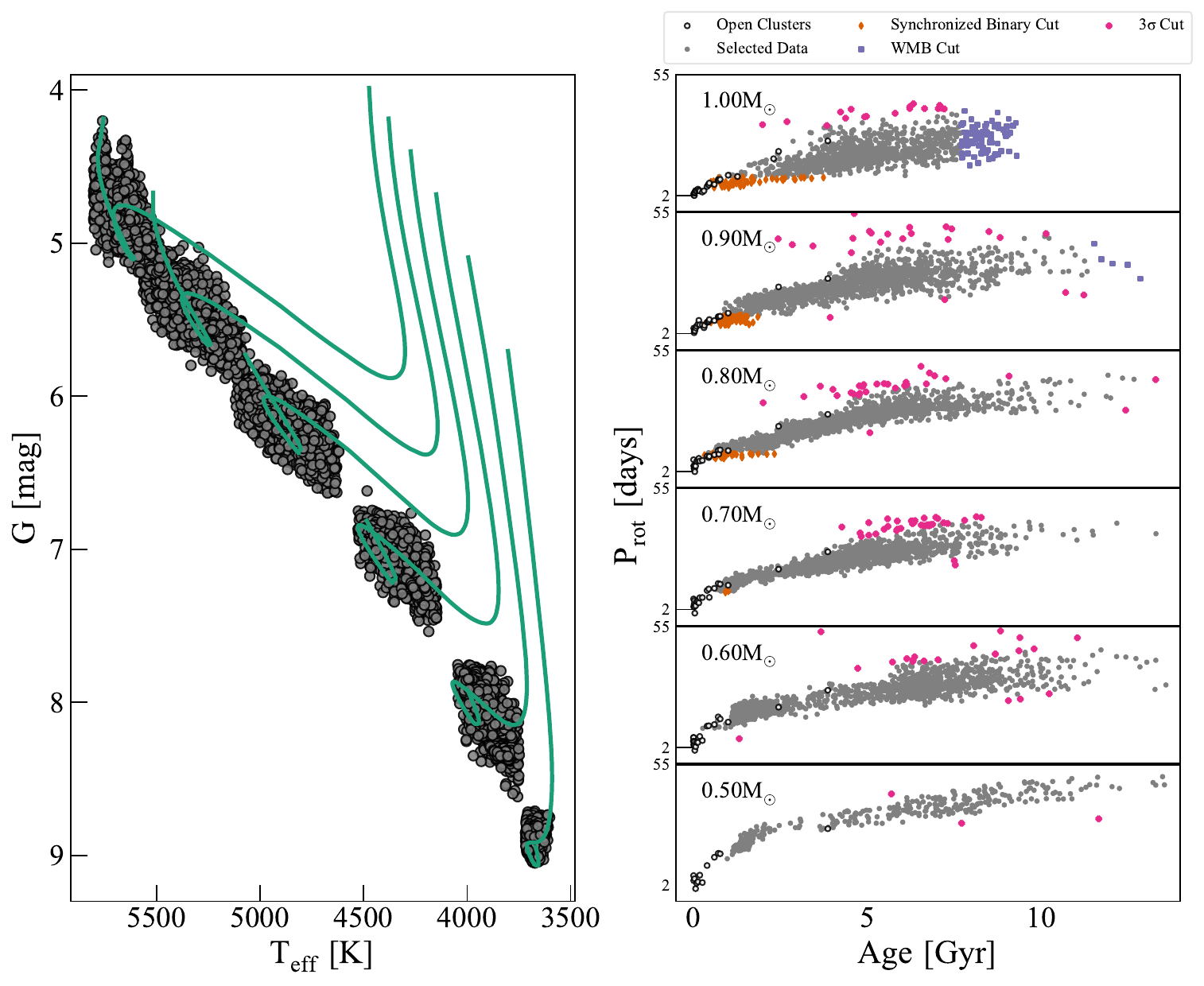}
    \caption{Data selection for mass-rotation sequences. Left: CMD data selection. In this panel, we show how stars are selected in the Gaia CMD based on proximity to the main sequence (colored solid lines) a given stellar evolution mass track (solid black lines). The selected stars are colored according to their stellar masses. Right: Rotation-age sequences for different stellar masses. In this panel we show the selected stars rotation-age distributions. These sequences have has 3 $\sigma$ outliers from the rolling median rotation period removed. In the lowest mass, one can see some evidence of core-envelope recouping at young ages and in the higher-masses, one can see the onset of weakened magnetic braking. }
    \label{fig: mass sequences}
\end{figure*}

After our sample selections, we have $7598$ gyro-kinematic stars to use in our model fits. Our selected sample and the corresponding mass sequences is given in Fig. \ref{fig: mass sequences}. For the full forward fits, we also exclude stars that are in the weakened magnetic braking regime with $Ro > 2.2$ according to the thresholds set by \citet{2016Natur.529..181V} and \citet{2025ApJ...986..120M}.

\subsection{Ensemble Sampler}

For each of our model cases and mass sequences, we use a MCMC sampler to determine the best fit model parameters. We set flat priors on the fitting parameters. Our likelihood function is given by: 

\begin{equation}\label{eq:likelihood}
    \log \mathcal{L} = -\frac{1}{2} \Sigma \bigg[\frac{(P_{rot, obs} - P_{rot, mod})^2}{\sigma P_{rot}^2}\bigg]  .
\end{equation}

Using the python package, \verb|emcee|, our MCMC sampler computes a burn in run with $100$ steps and $60$ walkers to get an initial estimate of the best fit parameters \citep{2013PASP..125..306F}. The result from the burn in run is then fed into the main run as an initial parameter distribution. The main fitting routine uses $60$ walkers taking $1000$ steps to determine the best fit parameters. The same likelihood function is applied to the analytic model fits, but uses $200$ walkers and $10000$ steps. The age range of the mass sequences used for the analytic fits is cut according to the age limits discussed in Sec. \ref{sec: wind dominated domain}. 

Our forward model fits determine the best fit physical parameters for the magnetized wind prescription. To translate these results into a format that is comparable to the analytic results, we compute the wind spin down index of the best fit model and the average best fit across all of out mass sequences according to Eq. \ref{eq:windparameters}. We reiterate that the wind spin down index is representative of the efficiency of stellar spin down if \textit{only} the stellar wind were driving angular momentum evolution. 

\section{Results} \label{sec: model results}

In this section we report the results for the determination of the wind-dominated age domain, the analytic model fitting and the forward model fitting. 

\subsection{Wind Dominated Domain}\label{sec: wind dominated domain}

We fit for the spin down slope in the wind-dominated regime, which excludes the youngest and oldest stars as outlined in Section \ref{sec: models} and summarized in Fig. \ref{fig: wind dominated region}. At lower masses, the treatment of internal angular momentum transport leaves a lasting imprint on the later rotational evolution of the star, up to $10$ Gyr. On the higher mass end, the contributions of evolutionary radius expansion to spin down, acts as the greatest limiting factor. Below a mass of $0.85 M_{\odot}$ radius expansion does not play an appreciable role. 

There has been a tendency in the literature to equate the empirical slope of the spin down relationship to the functional form of the wind torque alone. Our lower age limits suggest that empirical measurements of stellar spin down in low mass stars can be strongly influenced by structural evolution and the treatment of internal angular momentum transport mechanisms. The mass domain over which we expect empirical measurements of stellar spin down to be most closely related to the wind spin down index is between $\approx 0.7 - 0.9 M_{\odot}$. 

\subsection{Best fit Analytic Models}

\begin{figure*}
    \centering
    \includegraphics[width=\textwidth]{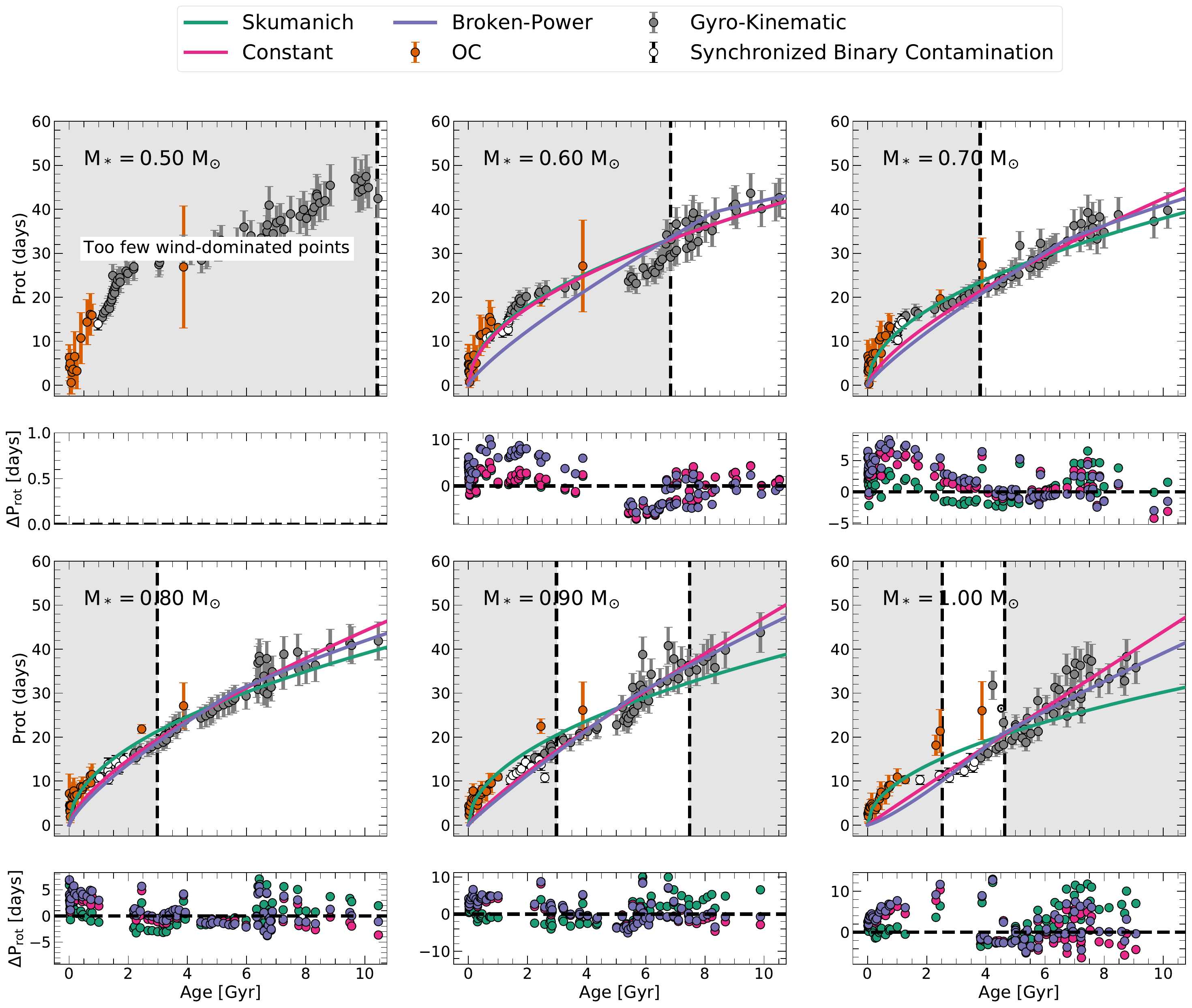}
    \caption{Analytic Model fits and residuals to all mass sequences. Each panel of this figure shows a $P_{rot}$ v. Age diagram for a given mass sequence. The grey points track the median rotation period of the sample and the orange points are open cluster measurements from the ChronoFlow training set. The white points are gyro-kinematic measurements, however they are below the synchronized binary contamination limit and are therefore excluded from our fits. Dashed black lines in the upper panels show the age limits imposed by assuming a wind-dominated domain. Shaded regions are excluded from the fitting process. Colored lines show the different analytic model fits discussed in section \ref{sec: analytic models}. The bottom panels of each row show the residuals between the observed rotation periods and the best fit analytic models. }
    \label{fig:analytic fits} 
    
\end{figure*}

The best fit analytic models are shown against the mass sequences in Fig. \ref{fig:analytic fits} and the corresponding best fit parameters are given in Fig. \ref{fig:best fit analytic}. Over the mass-range of $0.5-0.8 \ M_{\odot}$ there is little statistical significance to favor any one analytic models over the others according to a reduced $\chi^2$ statistic.

\subsubsection{Constant Spin Down}
If we average across all of the mass sequences to find the mean best fit spin down index from the constant spin down model we calculate a value of $0.748 \pm 0.089$ in the wind dominated domain. This is value is larger than the spin down index measured in older open clusters from the literature, but is still within statistical agreement \citep{2022ApJ...938..118D}. We measure a mean super-Skumanich spin down index of $0.746 \pm 0.053$ over the the wind-dominated domain in the mass range $0.70 - 0.90 \ M_{\odot}$ , which is an ideal laboratory for empirically measuring wind-driven spin down. 

\subsubsection{Weakened Magnetic Braking in the Gyro-Kinematic Sample}

\begin{figure}
    \centering
    \includegraphics[width=0.5\textwidth]{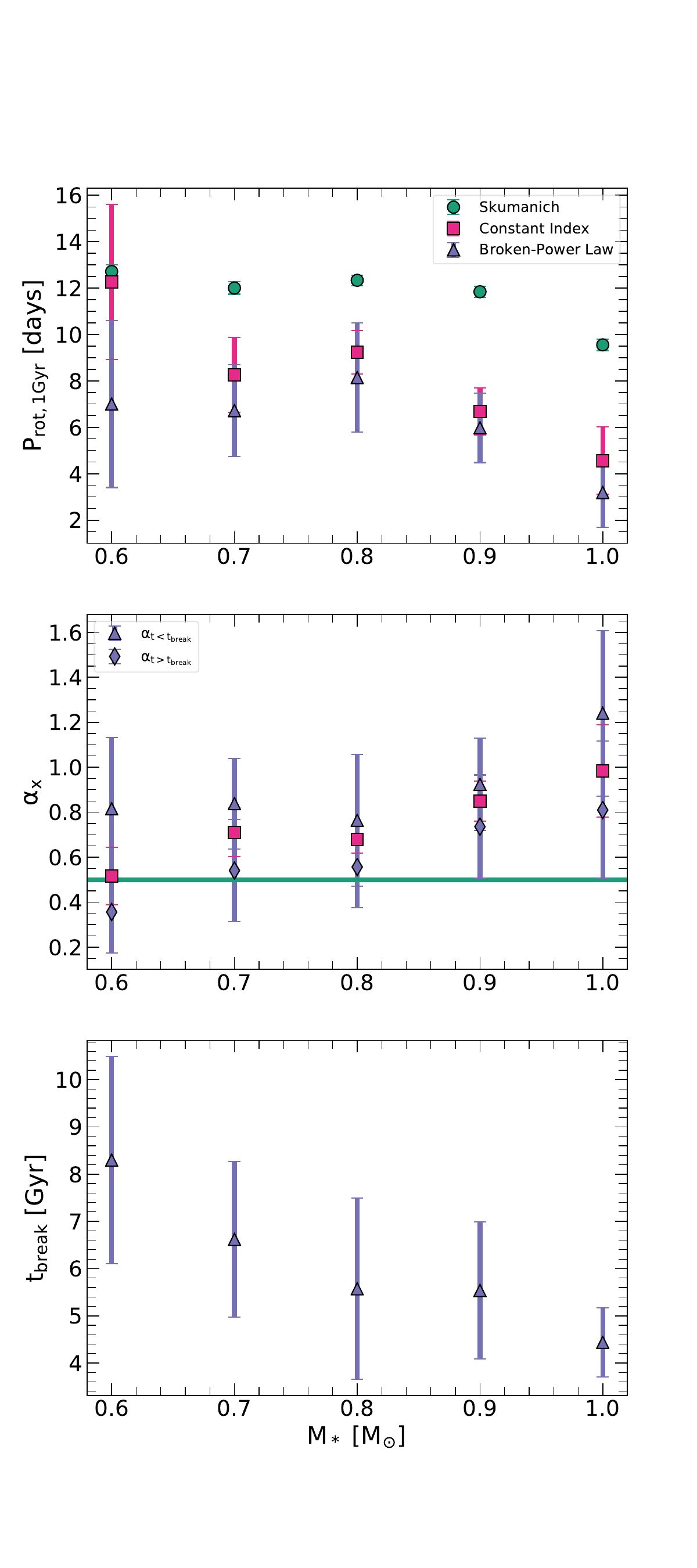}
    \caption{Best fits for parameters from Eq. \ref{eq:weakened magnetic braking}. Upper panel: This figure shows the best fit rotation period at 1 Gyr for each of the mass sequences. Middle panel: This panel shows how the best fit spin index of the constant spin model and the broken-power model change with mass. The purple points are the indices for the broken-power law model with the triangles representing the spin index before the onset of weakened magnetic braking, and the diamonds representing the post-braking index. The pink squares show the best fit spin index assuming a constant spin down. The teal line shows a Skumanich-like index. Bottom panel: Best fit braking age for different mass sequences.  }
    \label{fig:best fit analytic}  
\end{figure}

We present the best fit parameters for the broken-power law spin index fits in the bottom panel of Fig. \ref{fig:best fit analytic}. Above masses of $0.8 \ M_{\odot}$, the age where the power law breaks appears to decrease with increasing mass. This is consistent with the prediction that the onset of weakened magnetic braking occurs at a fixed Rossby number \citep{2016Natur.529..181V}.  The middle panel of Fig. \ref{fig:best fit analytic} shows the suppression in the spin down index before and after the onset of weakened magnetic braking. Generally we see a super-Skumanich value before the onset of weakened magnetic braking and a sub-Skumanich index following onset of weakened magnetic braking. The mean difference in the spin down index prior to the braking time and following it is $0.314$ across all of the masses. The best fit age for the onset of WMB can be translated to the Rossby number for the WMB transition. For the solar mass sequence, we find this Rossby threshold occurs at $1.79 \pm 1.01$, and at $1.95 \pm 0.74$ for  $0.9 M_{\odot}$. These are statistically consistent with literature measurements for the onset of WMB. Below this mass, we do not recover a signature of WMB, likely due to a selection effects. Additionally, if the threshold of WMB occurs at a constant Rossby number, then we expect that less stars will reach this stage at ages greater than $9$ Gyr which is the upper echelon of our sample. 

\subsection{Forward Model Fits}

\begin{figure*}
    \centering
    \includegraphics[width=\textwidth]{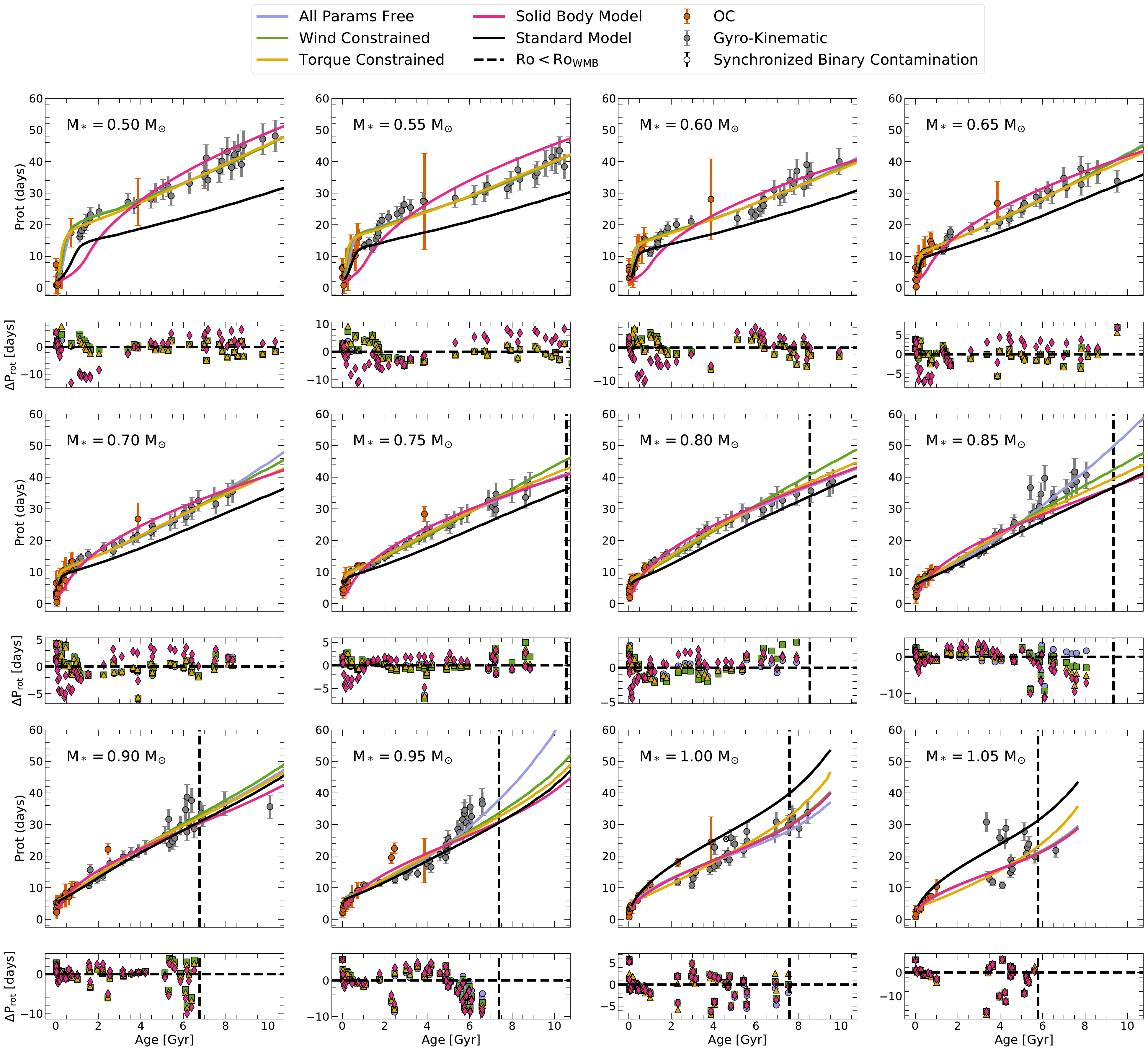}
    \caption{Model fits to individual mass sequences in $P_{rot}$ vs. Age space. Each panel includes data a different stellar mass increasing diagonally to the bottom right. The forward models are described in \ref{sec: models}. The best fit parameters for each of these model fits are plotted in \ref{fig:composite params} and the reduced $\chi^2$ values for each model fit are provided in Fig. \ref{fig:X2_forward}. Additionally, the ''standard" model case is given by the black curve in each panel. The vertical dashed lines show the predicted onset of weakened magnetic braking according to literature Rossby thresholds. }
    \label{fig:rotevol fits} 
    
\end{figure*}

\begin{figure*}
    \centering
    \includegraphics[width=0.95\textwidth]{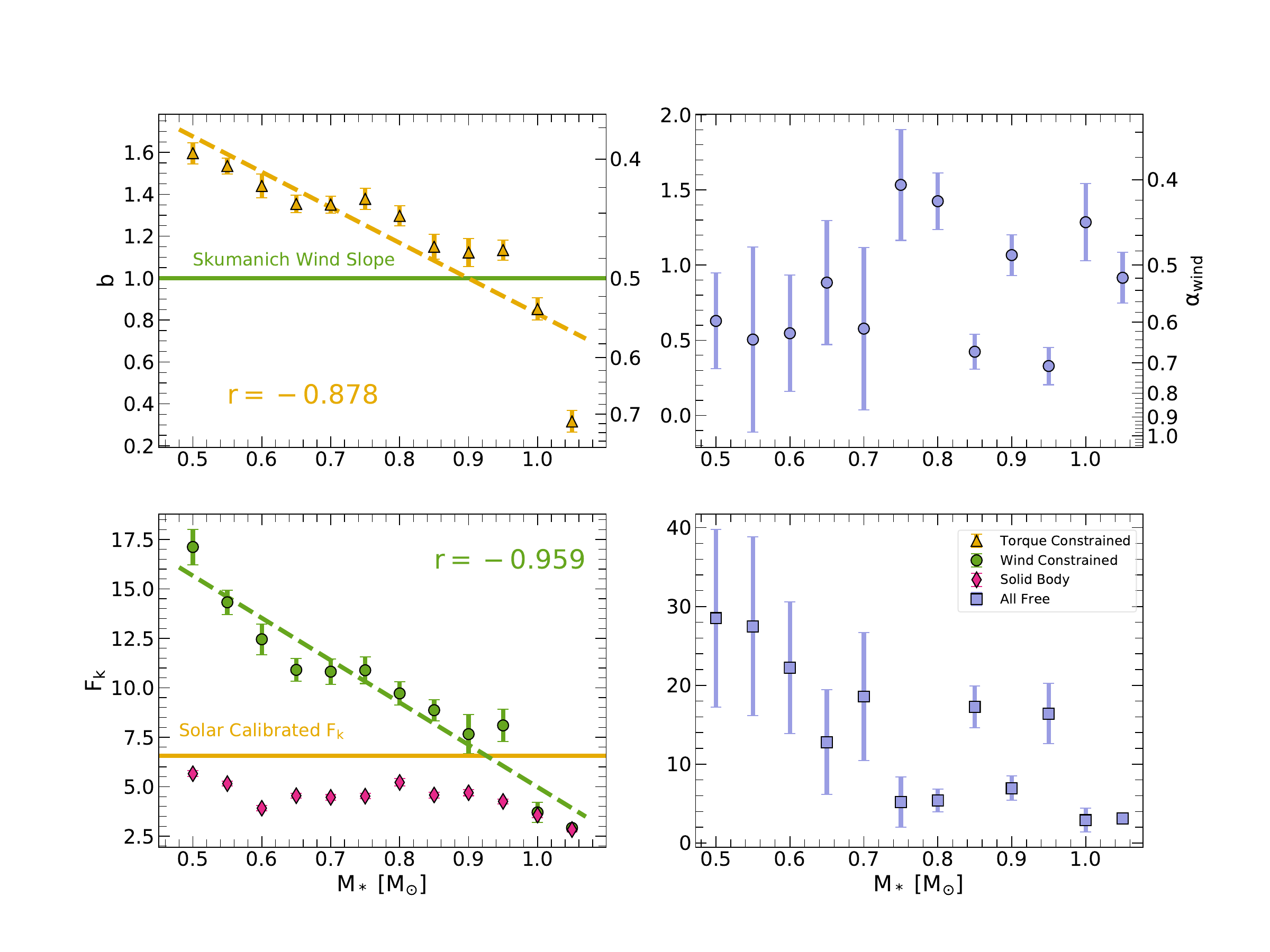}
    \caption{Best fit parameters for the forward models. Upper Left: Strength of the relationship between the magnetic field strength and angular velocity for the mass sequences. The right axis gives the corresponding wind spin down index. The solid green line is reproduces a Skumanich-like wind index. Upper right: Same as upper left panel, but for the least constrained model scenario. Lower left: Best fit torque scaling factor for different mass sequences. The yellow line gives the solar-calibrated torque, used in the torque constrained model. Lower right: Same as lower left, but for the least constrained model scenario. } 
    \label{fig:composite params}
\end{figure*}

\begin{figure}
    \centering
    \includegraphics[width=0.5\textwidth]{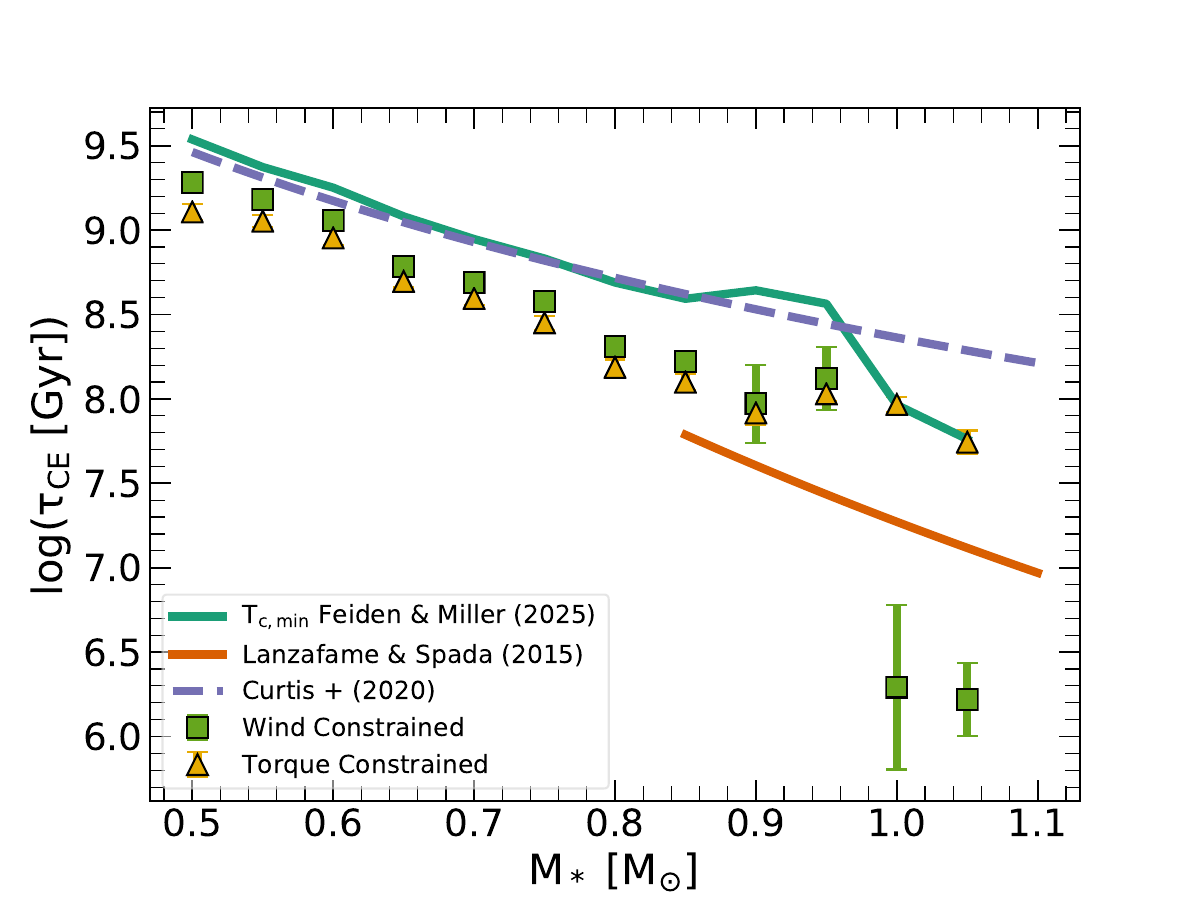}
    \caption{Core-envelope coupling timescale as a function of mass. The results are shown for the torque-constrained model and the wind-constrained model only as the model with all the parameters free is disfavored by the reduced $\chi^2$ statistic. The best fit coupling timescales are compared to prior literature estimates of stalling in solar-like stars (orange solid), the K dwarf regime (purple dashed), and the onset of a minimum central temperature (teal, solid) \citep{2010ApJ...716.1269D, 2020ApJ...904..140C, 2023RNAAS...7..161F}.} 
    \label{fig:fit wind index}
\end{figure}

\begin{figure}
    \centering
    \includegraphics[width=0.5\textwidth]{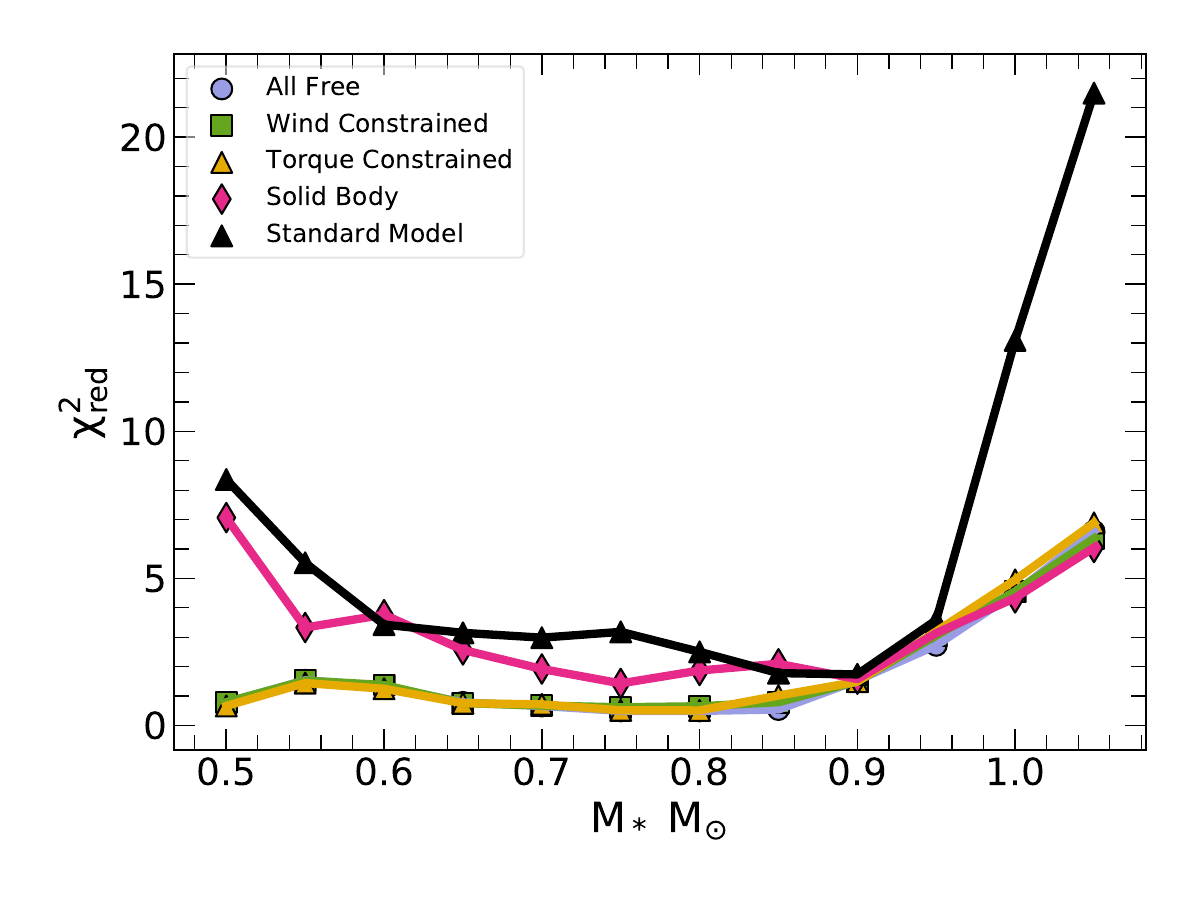}
    \caption{Reduced $\chi^2$ statistic for the forward model scenarios for each mass sequence. The black points give the Reduced $\chi^2$ value for a standard model which assumes a solar-calibrated torque scaling factor ans a Skumanich wind index.}
    \label{fig:X2_forward} 
\end{figure}

Our forward model fits find that a model which either allows for a variable torque scaling factor, $F_k$, or a changing wind-induced spin down slope best, moderated by $b$, is favored over a standard, solar-calibrated model approach. Additionally, in both cases we are able to recover a mass-dependent core-envelope timescale. The non-solid body models produce a more statistically consistent fit with the lower mass stars of our sample than a solid body model. Figure \ref{fig:rotevol fits} shows the model fits to the mass sequences and Figure \ref{fig:composite params} shows the best fit model parameters as a function of mass. These parameters are translated into a more useful wind spin down index in Fig. \ref{fig:fit wind index}. In Fig. \ref{fig:X2_forward} we compute the reduced $\chi^2$ statistic for each of our model scenarios across the mass sequences. 

The reduced $\chi^2$ statistic indicates that the gyro-kinematic data is equally well represented by a model which allows for either a free wind index \textit{or} a free torque scaling factor. To translate these to physical analog, a change in the torque scaling factor will change both the strength of the wind which will correspond to a change in the mass loss rate, the Alfvén radius, or some combination of the two. A change in the wind slope as a function of mass is a statement about an additional mass-dependence in the dynamo scalings. In other words, at different masses, the field strength or mass loss rate would have different dependencies on $\omega$. At lower masses, we see the torque scaling factor must be larger to adequately model the evolutionary sequences. This corresponds to the mass rate and Alfvén radius being larger than what is predicted by a solar-scaled model. This would require either the mass loss rate, field strength, or some combination to be greater than expected at a fixed scale. For the wind slope, we find low mass stars require a stronger relationship between the field strength or mass loss rate with angular velocity. Both of these effects are plausible degrees of freedom which deserve to be further vetted. 

If we compare a ''standard" wind model, to the mass-rotation sequences, we demonstrate the inability of such a model to be extrapolated to the old field star domain. At all masses, the reduced $\chi^2$ statistic disfavors this model. This standard model under-predicts rotation periods in the low mass domain of our sample, and over-predicts them in the high-mass domain.

\section{Discussion} \label{sec: discussion}

In this section, we explore the physical interpretations of our model fitting results. 

\subsection{Implications on Solar Scaled Wind Laws}

Our model forward model fits demonstrate the inability of a solar-scaled wind law to extrapolate to the old, slowly rotating domain. Our favored models indicate that either the Rossby scalings used to scale the mass loss rate and field strength do not capture all of the physical dependencies in the wind or there is an additional mass-dependence in the exponent of the Rossby scalings. For the torque-scaling model, this may be a statement about the determination of the Rossby number itself. This factor, $F_k$ is required to be stronger at lower masses to describe the late stage rotational behavior of these stars. The Rossby scaled wind model in its current form does not capture the observed mass dependence in spin down. The Rossby number in Eq. \ref{eq: Jdot} depends on how the convective overturn timescale is measured from models, so it is possible that the definition of convective overturn timescale is not fully capturing the relationship between the wind properties and the Rossby number. Finally, we rely on model structures (Eq. \ref{eq: structure} to compute the torques, therefore errors in quantities such as model radii or luminosities can propagate to errors in the predicted torques that are captured as a mass-dependent $F_k$ term. This interpretation of the departure from standard solar-scaled wind laws will be further explored in future work.

We caution that the kinematic ages may be subject to bias at old ages in for solar-mass stars. As the kinematic ages are inferred by grouping stars of similar properties together, they may be insensitive to rapid evolution at late ages on the MS. As solar-like stars near the MS turnoff point, they will experience rapid changes in their HR-diagram properties that are not fully encapsulated by the kinematic age method. This may explain why we see such a stark departure from the standard model in the $0.95 - 1.05 M_{\odot}$ samples. This bias induced by rapid evolution is not seen at lower masses as these stars have not yet reached the phase where they are approaching the subgiant branch. We therefore caution the over-interpretation of our results above $0.95 M_{\odot}$, both for the torque scaling results and the implications on the dynamo scalings discussed in Sec. \ref{sec: dynamo results}. 

An alternative way to cast the mass-dependence in the torque scaling factor is to hold the scaling factor constant and allow the mass-dependence to be absorbed by the Rossby number itself. That is, if the torque scaling factor is independent of mass, the Rossby number in its current form does not capture the rotational behavior of all stars at all ages.

\begin{figure*} \label{fig:gyrochrones}
    \centering
    \includegraphics[width=\textwidth]{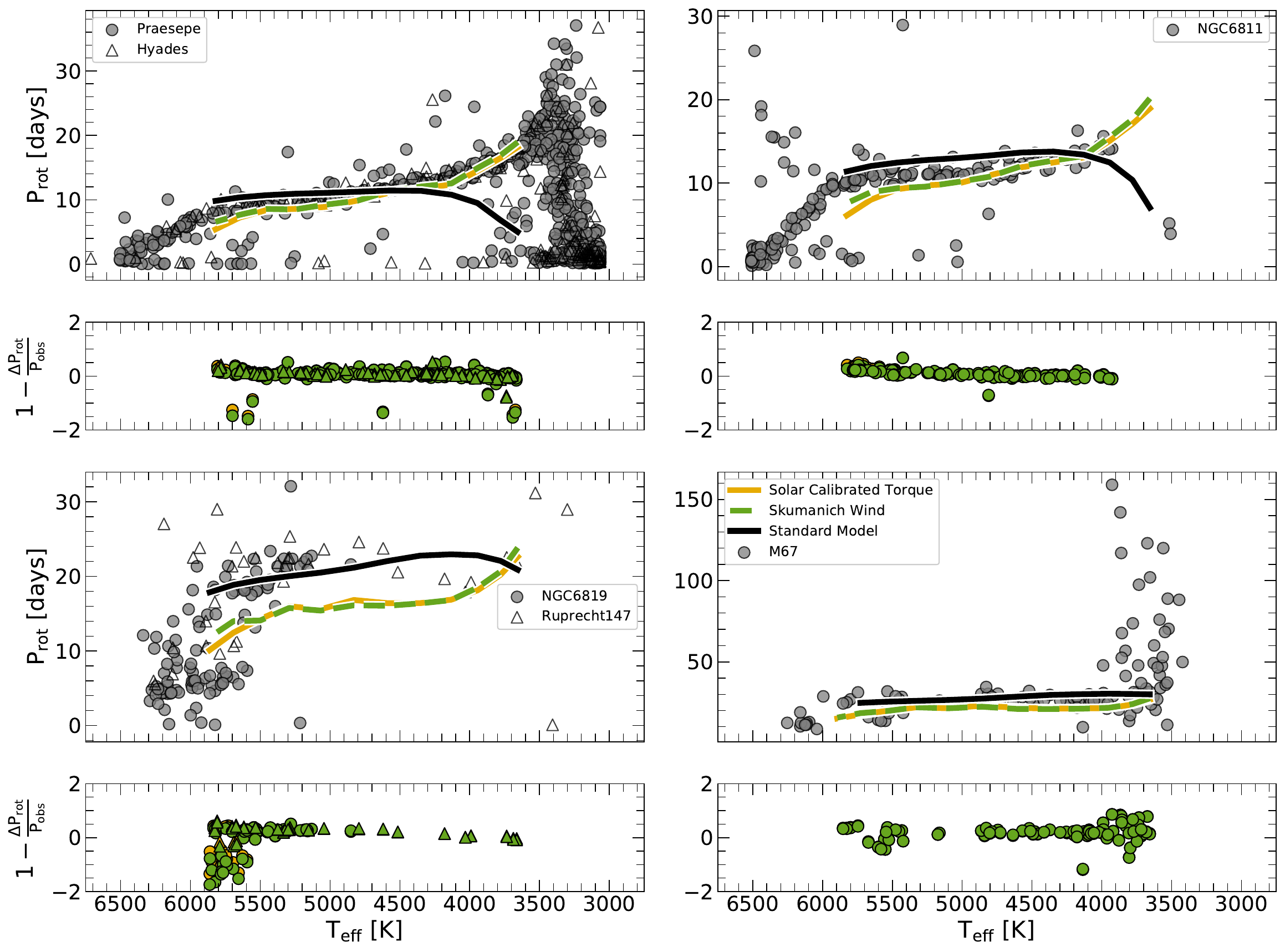}
    \caption{$P_{rot}$ v. $T_{eff}$ distributions for open clusters compared to our best fit models. Each panel shows different open clusters with similar ages along with the standard model predictions (black), the torque constrained model (yellow), and the wind constrained model (green). Open cluster data is taken from the \citet{2025ApJ...986...59V}.}
    
\end{figure*}

This result at face value appears to be in tension with results from the empirical spin down in open clusters, which  defines standard, "Skumanich" spin down and generally shows little evidence of a mass-dependent spin down index. As a final test of our modeling prescription, we build gyrochrones from our best fit model to compare against the cluster population and the gyro-kinematic data. These comparisons are made in Fig. \ref{fig:gyrochrones}. Our gyrochrones produce reasonable agreement with the intermediate age clusters in our sample. The exception to this is in NGC6819 and Ruprecht 147 where our model predicts faster rotation periods than anticipated. This age coincides with the stalling epoch noted in \citet{2020ApJ...904..140C}. There are several factors which may contribute to this disagreement including limitations in the core-envelope coupling prescription used in the model or metallicity effects. The exact source of this disagreement is beyond the scope of this paper. The agreement at other ages however is encouraging as we did not explicitly calibrate our model to reproduce the rotation period distribution at these younger ages. This also further demonstrates that the free parameter choices in our modeling approach has a greater effect on the predicted rotation periods at old ages, than at the young, cluster age scale. In the gyro-kinematic domain, we again recover fairly good agreement up until the oldest stars of our sample, near 10 Gyr. Our best fit models are over-predicting the rotation periods in the oldest stars of the sample. This is expected however, as shown by the analytic results, these stars are beyond the weakened magnetic braking threshold. As our forward model does not consider a mechanism by which to suppress spin down at the weakened magnetic braking threshold, therefore it is expected that the model will over-predict the rotation periods of these stars.

We propose that this departure from classic spin down has not been seen in open cluster measurements due to a combination of selection effects over which the spin-down index is measured in open clusters and an effect which becomes more apparent when looking at gyrochrones. To the former point, different literature studies measure different spin down rates from the open clusters \citep[e.g]{2015MNRAS.450.1787A, 2022ApJ...933..114D}. If the mass ranges over which the spin down index is measured do not match between studies, one would expect different values for the mean spin down rate. This is analogous to averaging different regions of the upper left panel of Fig. \ref{fig:composite params} to define a mass-independent index. Additionally, \citet{2026arXiv260325792P} show that gyrochrones, which use a full treatment of internal angular momentum transport and a single wind index across stellar masses, are unable to capture the full range of behavior for open clusters at different ages. In the youngest clusters, their models over predict the rotation periods of stars with near solar and greater than solar temperatures. It intermediate ages, the lower temperature cluster members rotate slower than what is predicted by the gyrochrones.

\subsection{Constraints on Dynamo Scalings} \label{sec: dynamo results}

By forward modeling the evolution of stars in the gyro-kinematic sample we can disentangle the effects of evolutionary radius expansion and the magnetized wind on angular momentum evolution. Perhaps the most interesting feature of our forward modeling results is that the gyro-kinematic data is equally well explained by a model where the magnetized wind properties are not consistent between stellar masses. As the structural effects are considered in our full forward model, this feature appears to be independent of inertial changes on the main sequence. 

Above we discussed a models where the scaling constant in the wind law, $F_k$, varies from predicted scalings. However, we have found another model solution which allows stars of different masses to share a common scaling factor, but allows the dynamo relationships to be mass-dependent. If the Rossby scalings are not to blame for this apparent departure from standard rotation models, then the slope of the dynamo scalings is. A change in this slope could source from the mass loss rate or the field strength's dependence on the angular velocity of the star. In standard wind prescriptions, one typically assumes that the relationship between the magnetic field strength and the angular velocity is linear, however our preferred model finds that the relationship decreases from a value near $1.6$ for $0.5 M_{\odot}$ stars to less than $0.5$ in our highest mass sequence. This corresponds to an increasing wind strength at higher masses, curiously consistent with the qualitative results from the analytic model fitting. As the forward model considers the stars radial changes on the main sequence, this enhancement in the spin index cannot be fully explained by evolutionary expansion. This feature has not been recorded in the literature, and is worthy of additional investigation and scrutiny. 

If the dynamo scalings of the torque scaling factor are the source of the apparent mass-dependence in the spin down relation, then independent data is needed to distinguish between the two potential contributors, the mass loss rate and the field strength. These constraints can be further probed by larger samples with more direct inferences of the stellar magnetic field. Zeeman Doppler Imaging is a promising method to measure the large scale magnetic field strengths and determine how they scale with the angular velocity of a star in the age domain of $1 - 10 \ Gyr$. By focusing efforts on the slowly rotating domain, we may be able to further disentangle the relationship between magnetic field generation and rotation. Other magnetic field indicators such as x-ray luminosities or H$\alpha$ emission may also help constrain this relationship, but some studies are indicate that these metrics probe smaller scale magnetic-fields and may not be as important in modulating the stellar rotation rate \citep{2024NatAs...8..223L}. 

\subsection{Recovery of Core-Envelope Coupling Timescales and Stalling}

For stars with masses $\le 0.85 M_{\odot}$, a non-solid body model is preferred over the solid body case, particularly for low mass stars with deep convective zones. While the signatures of core-envelope coupling are observed in the literature for higher mass stars, the relevant timescales are too short to show up in our sample making the solid body model and the models which consider non-solid body physics statistically indistinguishable from one another. This illustrates the necessity to consider a prescription for internal angular momentum transport in rotational evolution models. 

Additionally, we find that the predicted core-envelope coupling timescale is qualitatively consistent with the stalling timescale predicted by \citet{2020ApJ...904..140C} and with the minimum central temperature metric predicted by \citet{2023RNAAS...7..161F}. Both studies find a mass-dependent stalling timescale, but use different assumptions about the . In \citet{2020ApJ...904..140C} the mass-dependence goes as $M_*^{-3.65}$. After which, they assume that normal stellar spin down will resume. It is important to note that they use a super-Skumanich spin down index of $0.62$ from \citet{2019ApJ...879..100D} to fit for this timescale. They caution that this timescale should not be taken used as an absolute indicator of whether a star is stalled or not due to limitations in the braking index. Changes in the assumed spin down index will propagate to the inferred stalling timescale. It is therefore interesting that our coupling timescales are consistent with \citet{2020ApJ...904..140C} despite differences in the spin down indices. At these masses, our best fit spin down indices are sub-Skumanich, yet we still recover a consistent spin down timescale.  While our results are statistically consistent with literature predictions, the central values of our fits tend towards shorter coupling times than the literature predictions. The torque constrained model predicts a slightly longer coupling timescale than the wind-constrained model in this mass range. 

\begin{figure}
    \centering
    \includegraphics[width=0.5\textwidth]{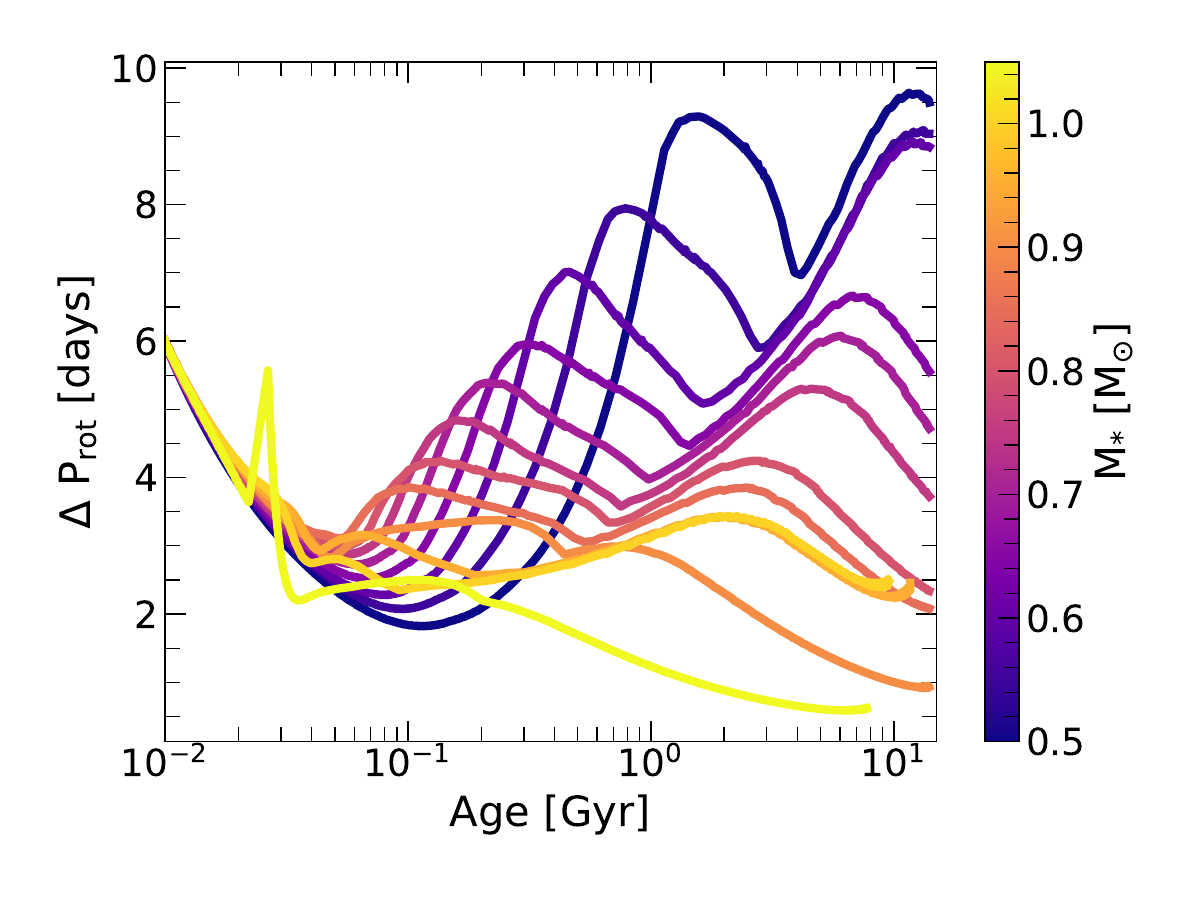}
    \caption{Model Predicted width of rotation period distribution. This figure shows the difference in rotation period as a function of age for non-solid body models launched at an initial rotation period of $2$ and $8$ days. This figure demonstrates that interior physics leaves a lasting impression on the late time rotational evolution of partially-convective stars.}
    \label{fig:core-envelops Prot width}
\end{figure}

Our models additionally allow us to construct a testable prediction to further test the physics of core-envelope decoupling. Fig. \ref{fig:core-envelops Prot width} shows the expected width of the rotation period distribution as a function of age for different stellar masses. These predictions are computed using the best fit core-envelope coupling timescale applied to models with initial periods of $2$ and $8$ days. If our core-envelope timescale recovery is correct, we would expect to see an age dependent distribution in rotation periods for stars of different masses. Kinematic ages will allow us to probe the older rotation distributions predicted by this metric. A potential complication to this is that core-envelope de-coupling may have an additional dependency on rotation rate. This feature is seen in empirical fits and theoretical models \citep{2013A&A...556A..36G, 2015A&A...577A..98G, 2016ApJ...829...32S}. Hydrodynamic models predict more a mass-dependence in the efficiency of coupling, but also find more efficient coupling for rapid rotators. Wave transport models however will only be mass dependent. A narrower range of rotation periods at late ages would favor a hybrid model, rather than one that depends only on mass. 

\subsection{The Onset of Weakened Magnetic Braking in the Gyro-Kinematic Sample}

The final feature of our results that we will highlight is the consistency of the best fit parameters for the broken-power law model with literature predictions for the onset of weakened magnetic braking. In Fig. \ref{fig:best fit analytic}, we show the age of the onset of weakened magnetic braking  and the spin-down indices before and after the predicted onset of weakened magnetic braking as a function of mass. Our results show that higher mass stars tend to have an earlier onset of weakened magnetic braking, with solar-like stars experiencing weakened magnetic braking near solar-age. Additionally, the best models show a suppression in the spin down index following the onset of weakened magnetic braking. This is in agreement with observations from \citet{2016Natur.529..181V} and \citet{2025ApJ...991L..17M}. We note that the onset ages for stars with masses less than $\approx 0.80 M_{\odot}$ are likely spurious as the reduced Chi-square statistics disfavor the broken-power law model and these stars do not reach the critical Rossby number predicted for the onset of weakened magnetic braking. 

From the best fit age for the onset of weakened magnetic braking, we can infer the critical Rossby number at which weakened magnetic braking occurs in the gyro-kinematic sample. The Rossby number is given by Eq. \ref{eq:Ro} and depends on $\tau_{conv}$, the convective turnover timescale. This convective turnover timescale depends on the depth of the convective zone, and is therefore a function of stellar mass and age. Interpolating within the YREC models we can determine the convective turnover timescale for a star with a given mass at the onset of weakened magnetic braking. It is important to note that because convective turnover time is not a direct observable, that it is subject to the uncertainties in the physics inherent to the evolution model. Omitting the predictions from stars with masses $\le 0.80 M_{\odot}$ as these stars are not expected to experience weakened magnetic braking on the MS, we find a critical Rossby number that is consistent with literature estimates. This bolsters the ability gyro-kinematic ages to measure second-order phenomena in the rotational-evolution of older field stars.

\section{Conclusion} \label{sec: conclusion}

In this study we have used gyro-kinematic ages to measure the efficacy of standard, solar scaled stellar spin down models as a function of mass for older partially-convective field stars. Kinematic ages provide an exciting new opportunity to calibrate gyrochronology to age scales greater than those accessible by benchmark open clusters. Using these ages, we build rotation-age sequences for stars of different mass in kinematic age. These mass sequences are photometrically consistent with a YREC stellar evolution track at a given stellar mass, allowing us to control for later evolutionary states that may contaminate our sample.

By fitting stars of different masses separately, we find that a standard Rossby scaling does not capture the full mass-dependence of stellar spin down as shown in Figs. \ref{fig:rotevol fits} and \ref{fig:composite params}. Departures from standard, Skumanich-like spin down have been noted by previous empirical work. \citet{2023ApJ...947L...3B} and \citet{2025ApJ...986...59V} use observations of open clusters to make predictive gyrochronology models. As these models are built from empirical measurements as opposed to a physical model, the rotation periods predicted from these models incorporate all of the contributing effects to rotation. As such, predictions from these models do not correspond to a case where only the magnetic wind is acting to spin down a star. 

On the theoretical side, \citet{2020A&A...636A..76S, 2026A&A...706A.262S} have sought to model departures from standard spin down by combining the effects of internal angular momentum transport with the stellar wind. In \citet{2026A&A...706A.262S}, they recover a steep mass-dependency in the coupling timescale which is qualitatively consistent with our recovered trend while allowing for a mass-dependence in the the wind torque scaling factor. While the qualitative results for the coupling timescales share a similar trend between our results and the \citet{2026A&A...706A.262S}, the trends in the torque scaling factors differ. This may speak to differences between the datasets used to constrain these parameters. \citet{2026A&A...706A.262S} uses open cluster data to a maximum age of a Gyr where as our results extend to $\approx 9 \ Gyr$ depending on the mass bin used. Other models have explored departures from stellar spin via the wind law itself. For instance, \citet{2017ApJ...845...46F, 2018ApJ...854...78F} determine how higher order magnetic field terms effect the stellar wind driven spin down of a star. They find two domains in which different field components will dominate the torque acting on the star. In the domain where the spin down is dominated by a quadrupolar field, the torque acting on the star will deviate from the more classic, dipolar dominated domain. 

In our study we account for departures from standard spin down with two best fit model families which show the viability of a mass-dependent term to replicate the spin-down of old, partially-convective stars. This mass-dependence can be explained either by the strength of the torque or in the dynamo relations. The first family of solutions requires that the torques in the least massive partially-convective stars are a factor of $\approx 2 $ greater than a solar-calibrated model. The change in the torque could be capturing defects in the Rossby number or model predictions themselves. A correction term to the convective turnover timescale and therefore the Rossby number could plausibly explain this effect. Additionally, there are well-known issues in model predictions of lower mass MS stars. A more complete physical treatment of the input structure models for cool stars, including effects such as starspot driven radius expansion or a more physically motivated model atmospheres may alleviate some of the mass-dependence in the torque scaling factor. Both of these potential explanations will be explored in future work. 

An equally viable solution to the mass-dependent deviations in spin down is that the dynamo scalings have a mass dependence. We demonstrate that constant, Skumanich-like index of $0.5$ fails to reproduce observations. Rather, lower mass stars in our sample require a weaker time dependence in the period evolution. This change in the overall wind spin down index could be accounted for by either a mass-dependence between the mass loss rate and the angular velocity, a mass-dependence between the global field strength and angular velocity, or some combination of the two. Exploring field strength metrics measured from techniques such as Zeeman-Doppler imaging or polarimetry for stars of different masses and measuring the slope of these metrics with angular velocity is a promising avenue for breaking the degeneracy between our mass-dependent dynamo model and our mass-dependent torque model.  

Our analytic and physical model fits also point to implications for the interpretation of empirical spin down measurements. These implications are described below: 

\begin{enumerate}
    \item In combination with the inferences of the wind-dominated domain, our forward modeling demonstrates that one cannot extrapolate a simple Skumanich-like gyrochronological relationships to the old stellar domain. More specifically, we demonstrate one must consider structural effects such as core-envelope de-coupling and evolutionary radius expansion on the main sequence to adequately explain the rotation periods of old, partially-convective stars. In future work calibrating gyrochronology, we advocate for the use of forward models which consider structural effects alongside the magnetized stellar wind. 
    \item Core-envelope de-coupling has a strikingly strong impact on the predicted rotation evolution, even well beyond the stalling domain. We recover a qualitatively consistent trend in the best fit core-envelope coupling timescale and literature studies of stalling in partially-convective stars. This mass-dependent timescale can be further vetted using model predictions of the width of rotation period distributions as a function of age. 
    \item The best fit parameters from our broken-power law analytic model are consistent with predictions for the onset of weakened magnetic braking predicted by \citet{2016Natur.529..181V}. This demonstrates the ability of kinematic ages to probe the physics of spin down in this older age domain. 
\end{enumerate}

We demonstrate the complex nature of defining a wind-dominated domain, which in turn demonstrates the limited domain over-which simple power-law gyrochronology can be applied. We argue that the effects of internal angular-momentum distribution and evolutionary inertial changes are critical for interpreting the spin down of partially-convective starts at young and old ages respectively.

We note that this wind prescription used in this study is one of many in the literature. In future work this data may be used to constrain the allowed parameter space of other models such as those by \citet{2018ApJ...864..125F}. These models, informed by results from MHD simulations explore the effect of variability on the magnetized winds. Additionally, it may be fruitful to explore wind models which consider higher order magnetic field contributions such as those presented in \citet{2018MNRAS.474..536S} and \citet{2019ApJ...886..120S} using a combination of gyro-kinematic ages and magnetic field strength metrics. 

Future work will aim to address questions in how the stellar spin down properties measured in this study change with metallicity and with stellar structure. Specifically, the techniques presented in this work can be applied to the fully-convective domain where a number of interesting contradictions about rotational and magnetic evolution remain unsolved. Additionally, this type of analysis would benefit from a more sophisticated modeling approach which uses considers structural effects due to spots and rotation as well as a prescription which self-consistently evolves rotation and starspots in tandem.

\section*{Acknowledgments               }

AA and MHP acknowledge support from NASA grant 80NSSC24K0622. MHP also acknowledges support from NSF grant 2507891. 

AA wishes to acknowledge the Ohio State stars research group for helpful input and guidance on the contents of this manuscript. AA also wishes to thank T. Thompson, J. van Saders and V. See for advice and insight into various physical components of this manuscript.

This work made use of Astropy:\footnote{http://www.astropy.org} a community-developed core Python package and an ecosystem of tools and resources for astronomy \citep{astropy:2013, astropy:2018, astropy:2022}.

\section*{Data Availability}

The gyro-kinematic age catalog used in this work was compiled using methods from \citet{2021AJ....161..189L, 2022AJ....164..251L}. The stellar evolution models in this work were computed using the Yale Rotating Evolution Code and complimentary code, Rotevol. The namelists and outputs for these models can be found at https://doi.org/10.5281/zenodo.21340200.


\bibliography{main}{}

@ARTICLE{2021AJ....161..189L,
       author = {{Lu}, Yuxi Lucy and {Angus}, Ruth and {Curtis}, Jason L. and {David}, Trevor J. and {Kiman}, Rocio},
        title = "{Gyro-kinematic Ages for around 30,000 Kepler Stars}",
      journal = {\aj},
         year = 2021,
        month = apr,
       volume = {161},
       number = {4},
          eid = {189},
        pages = {189},
          doi = {10.3847/1538-3881/abe4d6},
archivePrefix = {arXiv},
       eprint = {2102.01772},
 primaryClass = {astro-ph.SR},
       adsurl = {https://ui.adsabs.harvard.edu/abs/2021AJ....161..189L}
}

@ARTICLE{2016Natur.529..181V,
       author = {{van Saders}, Jennifer L. and {Ceillier}, Tugdual and {Metcalfe}, Travis S. and {Silva Aguirre}, Victor and {Pinsonneault}, Marc H. and {Garc{\'\i}a}, Rafael A. and {Mathur}, Savita and {Davies}, Guy R.},
        title = "{Weakened magnetic braking as the origin of anomalously rapid rotation in old field stars}",
      journal = {\nat},
         year = 2016,
        month = jan,
       volume = {529},
       number = {7585},
        pages = {181-184},
          doi = {10.1038/nature16168},
archivePrefix = {arXiv},
       eprint = {1601.02631},
 primaryClass = {astro-ph.SR},
       adsurl = {https://ui.adsabs.harvard.edu/abs/2016Natur.529..181V}
}

@ARTICLE{1972ApJ...171..565S,
       author = {{Skumanich}, A.},
        title = "{Time Scales for Ca II Emission Decay, Rotational Braking, and Lithium Depletion}",
      journal = {\apj},
         year = 1972,
        month = feb,
       volume = {171},
        pages = {565},
          doi = {10.1086/151310},
       adsurl = {https://ui.adsabs.harvard.edu/abs/1972ApJ...171..565S}
}

@ARTICLE{2020ApJ...904..140C,
       author = {{Curtis}, Jason Lee and {Ag{\"u}eros}, Marcel A. and {Matt}, Sean P. and {Covey}, Kevin R. and {Douglas}, Stephanie T. and {Angus}, Ruth and {Saar}, Steven H. and {Cody}, Ann Marie and {Vanderburg}, Andrew and {Law}, Nicholas M. and {Kraus}, Adam L. and {Latham}, David W. and {Baranec}, Christoph and {Riddle}, Reed and {Ziegler}, Carl and {Lund}, Mikkel N. and {Torres}, Guillermo and {Meibom}, S{\o}ren and {Aguirre}, Victor Silva and {Wright}, Jason T.},
        title = "{When Do Stalled Stars Resume Spinning Down? Advancing Gyrochronology with Ruprecht 147}",
      journal = {\apj},
         year = 2020,
        month = dec,
       volume = {904},
       number = {2},
          eid = {140},
        pages = {140},
          doi = {10.3847/1538-4357/abbf58},
archivePrefix = {arXiv},
       eprint = {2010.02272},
 primaryClass = {astro-ph.SR},
       adsurl = {https://ui.adsabs.harvard.edu/abs/2020ApJ...904..140C}
}

@ARTICLE{2022ApJ...938..118D,
       author = {{Dungee}, Ryan and {van Saders}, Jennifer and {Gaidos}, Eric and {Chun}, Mark and {Garc{\'\i}a}, Rafael A. and {Magnier}, Eugene A. and {Mathur}, Savita and {Santos}, {\^A}ngela R.~G.},
        title = "{A 4 Gyr M-dwarf Gyrochrone from CFHT/MegaPrime Monitoring of the Open Cluster M67}",
      journal = {\apj},
         year = 2022,
        month = oct,
       volume = {938},
       number = {2},
          eid = {118},
        pages = {118},
          doi = {10.3847/1538-4357/ac90be},
archivePrefix = {arXiv},
       eprint = {2211.01377},
 primaryClass = {astro-ph.SR},
       adsurl = {https://ui.adsabs.harvard.edu/abs/2022ApJ...938..118D}
}

@ARTICLE{2024AJ....167..159L,
       author = {{Lu}, Yuxi(Lucy) and {Angus}, Ruth and {Foreman-Mackey}, Daniel and {Hattori}, Soichiro},
        title = "{In This Day and Age: An Empirical Gyrochronology Relation for Partially and Fully Convective Single Field Stars}",
      journal = {\aj},
         year = 2024,
        month = apr,
       volume = {167},
       number = {4},
          eid = {159},
        pages = {159},
          doi = {10.3847/1538-3881/ad28b9},
archivePrefix = {arXiv},
       eprint = {2310.14990},
 primaryClass = {astro-ph.SR},
       adsurl = {https://ui.adsabs.harvard.edu/abs/2024AJ....167..159L}
}

@ARTICLE{2007ApJ...669.1167B,
       author = {{Barnes}, Sydney A.},
        title = "{Ages for Illustrative Field Stars Using Gyrochronology: Viability, Limitations, and Errors}",
      journal = {\apj},
         year = 2007,
        month = nov,
       volume = {669},
       number = {2},
        pages = {1167-1189},
          doi = {10.1086/519295},
archivePrefix = {arXiv},
       eprint = {0704.3068},
 primaryClass = {astro-ph},
       adsurl = {https://ui.adsabs.harvard.edu/abs/2007ApJ...669.1167B}
}

@ARTICLE{2024NatAs...8..223L,
       author = {{Lu}, Yuxi Lucy and {See}, Victor and {Amard}, Louis and {Angus}, Ruth and {Matt}, Sean P.},
        title = "{An abrupt change in the stellar spin-down law at the fully convective boundary}",
      journal = {Nature Astronomy},
         year = 2024,
        month = feb,
       volume = {8},
        pages = {223-229},
          doi = {10.1038/s41550-023-02126-2},
archivePrefix = {arXiv},
       eprint = {2306.09119},
 primaryClass = {astro-ph.SR},
       adsurl = {https://ui.adsabs.harvard.edu/abs/2024NatAs...8..223L}
}

@ARTICLE{2023A&A...674A..26C,
       author = {{Creevey}, O.~L. and {Sordo}, R. and {Pailler}, F. and {Fr{\'e}mat}, Y. and {Heiter}, U. and {Th{\'e}venin}, F. and {Andrae}, R. and {Fouesneau}, M. and {Lobel}, A. and {Bailer-Jones}, C.~A.~L. and {Garabato}, D. and {Bellas-Velidis}, I. and {Brugaletta}, E. and {Lorca}, A. and {Ordenovic}, C. and {Palicio}, P.~A. and {Sarro}, L.~M. and {Delchambre}, L. and {Drimmel}, R. and {Rybizki}, J. and {Torralba Elipe}, G. and {Korn}, A.~J. and {Recio-Blanco}, A. and {Schultheis}, M.~S. and {De Angeli}, F. and {Montegriffo}, P. and {Abreu Aramburu}, A. and {Accart}, S. and {{\'A}lvarez}, M.~A. and {Bakker}, J. and {Brouillet}, N. and {Burlacu}, A. and {Carballo}, R. and {Casamiquela}, L. and {Chiavassa}, A. and {Contursi}, G. and {Cooper}, W.~J. and {Dafonte}, C. and {Dapergolas}, A. and {de Laverny}, P. and {Dharmawardena}, T.~E. and {Edvardsson}, B. and {Le Fustec}, Y. and {Garc{\'\i}a-Lario}, P. and {Garc{\'\i}a-Torres}, M. and {Gomez}, A. and {Gonz{\'a}lez-Santamar{\'\i}a}, I. and {Hatzidimitriou}, D. and {Jean-Antoine Piccolo}, A. and {Kontiza}, M. and {Kordopatis}, G. and {Lanzafame}, A.~C. and {Lebreton}, Y. and {Licata}, E.~L. and {Lindstr{\o}m}, H.~E.~P. and {Livanou}, E. and {Magdaleno Romeo}, A. and {Manteiga}, M. and {Marocco}, F. and {Marshall}, D.~J. and {Mary}, N. and {Nicolas}, C. and {Pallas-Quintela}, L. and {Panem}, C. and {Pichon}, B. and {Poggio}, E. and {Riclet}, F. and {Robin}, C. and {Santove{\~n}a}, R. and {Silvelo}, A. and {Slezak}, I. and {Smart}, R.~L. and {Soubiran}, C. and {S{\"u}veges}, M. and {Ulla}, A. and {Utrilla}, E. and {Vallenari}, A. and {Zhao}, H. and {Zorec}, J. and {Barrado}, D. and {Bijaoui}, A. and {Bouret}, J. -C. and {Blomme}, R. and {Brott}, I. and {Cassisi}, S. and {Kochukhov}, O. and {Martayan}, C. and {Shulyak}, D. and {Silvester}, J.},
        title = "{Gaia Data Release 3. Astrophysical parameters inference system (Apsis). I. Methods and content overview}",
      journal = {\aap},
         year = 2023,
        month = jun,
       volume = {674},
          eid = {A26},
        pages = {A26},
          doi = {10.1051/0004-6361/202243688},
archivePrefix = {arXiv},
       eprint = {2206.05864},
 primaryClass = {astro-ph.GA},
       adsurl = {https://ui.adsabs.harvard.edu/abs/2023A&A...674A..26C}
}

@ARTICLE{2025ApJ...984..125L,
       author = {{Li}, Yaguang and {Huber}, Daniel and {Ong}, J.~M. Joel and {van Saders}, Jennifer and {Costa}, R.~R. and {Larsen}, Jens Reersted and {Basu}, Sarbani and {Bedding}, Timothy R. and {Dai}, Fei and {Chontos}, Ashley and {Carmichael}, Theron W. and {Hey}, Daniel and {Kjeldsen}, Hans and {Hon}, Marc and {Campante}, Tiago L. and {Monteiro}, M{\'a}rio J.~P.~F.~G. and {Lundkvist}, Mia Sloth and {Saunders}, Nicholas and {Isaacson}, Howard and {Howard}, Andrew W. and {Gibson}, Steven R. and {Halverson}, Samuel and {Rider}, Kodi and {Roy}, Arpita and {Baker}, Ashley D. and {Edelstein}, Jerry and {Smith}, Chris and {Fulton}, Benjamin J. and {Walawender}, Josh},
        title = "{K Dwarf Radius Inflation and a 10 Gyr Spin-down Clock Unveiled through Asteroseismology of HD 219134 from the Keck Planet Finder}",
      journal = {\apj},
         year = 2025,
        month = may,
       volume = {984},
       number = {2},
          eid = {125},
        pages = {125},
          doi = {10.3847/1538-4357/adc737},
archivePrefix = {arXiv},
       eprint = {2502.00971},
 primaryClass = {astro-ph.SR},
       adsurl = {https://ui.adsabs.harvard.edu/abs/2025ApJ...984..125L}
}

@ARTICLE{2015MNRAS.450.1787A,
       author = {{Angus}, Ruth and {Aigrain}, Suzanne and {Foreman-Mackey}, Daniel and {McQuillan}, Amy},
        title = "{Calibrating gyrochronology using Kepler asteroseismic targets}",
      journal = {\mnras},
         year = 2015,
        month = jun,
       volume = {450},
       number = {2},
        pages = {1787-1798},
          doi = {10.1093/mnras/stv423},
archivePrefix = {arXiv},
       eprint = {1502.06965},
 primaryClass = {astro-ph.EP},
       adsurl = {https://ui.adsabs.harvard.edu/abs/2015MNRAS.450.1787A}
}

@ARTICLE{2013A&A...556A..36G,
       author = {{Gallet}, F. and {Bouvier}, J.},
        title = "{Improved angular momentum evolution model for solar-like stars}",
      journal = {\aap},
         year = 2013,
        month = aug,
       volume = {556},
          eid = {A36},
        pages = {A36},
          doi = {10.1051/0004-6361/201321302},
archivePrefix = {arXiv},
       eprint = {1306.2130},
 primaryClass = {astro-ph.SR},
       adsurl = {https://ui.adsabs.harvard.edu/abs/2013A&A...556A..36G}
}

@ARTICLE{2023ApJS..267....8A,
       author = {{Andrae}, Ren{\'e} and {Rix}, Hans-Walter and {Chandra}, Vedant},
        title = "{Robust Data-driven Metallicities for 175 Million Stars from Gaia XP Spectra}",
      journal = {\apjs},
         year = 2023,
        month = jul,
       volume = {267},
       number = {1},
          eid = {8},
        pages = {8},
          doi = {10.3847/1538-4365/acd53e},
archivePrefix = {arXiv},
       eprint = {2302.02611},
 primaryClass = {astro-ph.SR},
       adsurl = {https://ui.adsabs.harvard.edu/abs/2023ApJS..267....8A}
}

@ARTICLE{2012ApJ...754L..26M,
       author = {{Matt}, Sean P. and {MacGregor}, Keith B. and {Pinsonneault}, Marc H. and {Greene}, Thomas P.},
        title = "{Magnetic Braking Formulation for Sun-like Stars: Dependence on Dipole Field Strength and Rotation Rate}",
      journal = {\apjl},
         year = 2012,
        month = aug,
       volume = {754},
       number = {2},
          eid = {L26},
        pages = {L26},
          doi = {10.1088/2041-8205/754/2/L26},
archivePrefix = {arXiv},
       eprint = {1206.2354},
 primaryClass = {astro-ph.SR},
       adsurl = {https://ui.adsabs.harvard.edu/abs/2012ApJ...754L..26M}
}

@ARTICLE{2024RNAAS...8..260M,
       author = {{Metcalfe}, Travis S. and {Corsaro}, Enrico and {Bonanno}, Alfio and {Creevey}, Orlagh L. and {van Saders}, Jennifer L.},
        title = "{Extending the Asteroseismic Calibration of the Stellar Rossby Number}",
      journal = {Research Notes of the American Astronomical Society},
         year = 2024,
        month = oct,
       volume = {8},
       number = {10},
          eid = {260},
        pages = {260},
          doi = {10.3847/2515-5172/ad8566},
archivePrefix = {arXiv},
       eprint = {2410.08106},
 primaryClass = {astro-ph.SR},
       adsurl = {https://ui.adsabs.harvard.edu/abs/2024RNAAS...8..260M}
}

@ARTICLE{2016ApJ...829...32S,
       author = {{Somers}, Garrett and {Pinsonneault}, Marc H.},
        title = "{Lithium Depletion is a Strong Test of Core-envelope Recoupling}",
      journal = {\apj},
         year = 2016,
        month = sep,
       volume = {829},
       number = {1},
          eid = {32},
        pages = {32},
          doi = {10.3847/0004-637X/829/1/32},
archivePrefix = {arXiv},
       eprint = {1606.00004},
 primaryClass = {astro-ph.SR},
       adsurl = {https://ui.adsabs.harvard.edu/abs/2016ApJ...829...32S}
}

@ARTICLE{2013ApJ...776...67V,
       author = {{van Saders}, Jennifer L. and {Pinsonneault}, Marc H.},
        title = "{Fast Star, Slow Star; Old Star, Young Star: Subgiant Rotation as a Population and Stellar Physics Diagnostic}",
      journal = {\apj},
         year = 2013,
        month = oct,
       volume = {776},
       number = {2},
          eid = {67},
        pages = {67},
          doi = {10.1088/0004-637X/776/2/67},
archivePrefix = {arXiv},
       eprint = {1306.3701},
 primaryClass = {astro-ph.SR},
       adsurl = {https://ui.adsabs.harvard.edu/abs/2013ApJ...776...67V}
}

@ARTICLE{2010ApJ...716.1269D,
       author = {{Denissenkov}, Pavel A. and {Pinsonneault}, Marc and {Terndrup}, Donald M. and {Newsham}, Grant},
        title = "{Angular Momentum Transport in Solar-type Stars: Testing the Timescale for Core-Envelope Coupling}",
      journal = {\apj},
         year = 2010,
        month = jun,
       volume = {716},
       number = {2},
        pages = {1269-1287},
          doi = {10.1088/0004-637X/716/2/1269},
archivePrefix = {arXiv},
       eprint = {0911.1121},
 primaryClass = {astro-ph.SR},
       adsurl = {https://ui.adsabs.harvard.edu/abs/2010ApJ...716.1269D}
}

@ARTICLE{2022AJ....164..251L,
       author = {{Lu}, Yuxi Lucy and {Curtis}, Jason L. and {Angus}, Ruth and {David}, Trevor J. and {Hattori}, Soichiro},
        title = "{Bridging the Gap-The Disappearance of the Intermediate Period Gap for Fully Convective Stars, Uncovered by New ZTF Rotation Periods}",
      journal = {\aj},
         year = 2022,
        month = dec,
       volume = {164},
       number = {6},
          eid = {251},
        pages = {251},
          doi = {10.3847/1538-3881/ac9bee},
archivePrefix = {arXiv},
       eprint = {2210.06604},
 primaryClass = {astro-ph.SR},
       adsurl = {https://ui.adsabs.harvard.edu/abs/2022AJ....164..251L}
}

@ARTICLE{2025ApJ...991L..17M,
       author = {{Metcalfe}, Travis S. and {van Saders}, Jennifer L. and {Pinsonneault}, Marc H. and {Ayres}, Thomas R. and {Kochukhov}, Oleg and {Stassun}, Keivan G. and {Finley}, Adam J. and {See}, Victor and {Ilyin}, Ilya V. and {Strassmeier}, Klaus G.},
        title = "{Weakened Magnetic Braking Signals the Collapse of the Global Stellar Dynamo}",
      journal = {\apjl},
         year = 2025,
        month = sep,
       volume = {991},
       number = {1},
          eid = {L17},
        pages = {L17},
          doi = {10.3847/2041-8213/ae03bc},
archivePrefix = {arXiv},
       eprint = {2509.03717},
 primaryClass = {astro-ph.SR},
       adsurl = {https://ui.adsabs.harvard.edu/abs/2025ApJ...991L..17M}
}

@ARTICLE{2025ApJ...986..120M,
       author = {{Metcalfe}, Travis S. and {Petit}, Pascal and {van Saders}, Jennifer L. and {Ayres}, Thomas R. and {Buzasi}, Derek and {Kochukhov}, Oleg and {Stassun}, Keivan G. and {Pinsonneault}, Marc H. and {Ilyin}, Ilya V. and {Strassmeier}, Klaus G. and {Finley}, Adam J. and {Garc{\'\i}a}, Rafael A. and {Huber}, Daniel and {Lu}, Yuxi (Lucy) and {See}, Victor},
        title = "{Testing the Rossby Paradigm: Weakened Magnetic Braking in Early K-type Stars}",
      journal = {\apj},
         year = 2025,
        month = jun,
       volume = {986},
       number = {2},
          eid = {120},
        pages = {120},
          doi = {10.3847/1538-4357/add40a},
archivePrefix = {arXiv},
       eprint = {2501.19169},
 primaryClass = {astro-ph.SR},
       adsurl = {https://ui.adsabs.harvard.edu/abs/2025ApJ...986..120M}
}

@ARTICLE{2023RNAAS...7..161F,
       author = {{Feiden}, Gregory A. and {Miller}, Lydia},
        title = "{Rotational Stalling Ends when Main-sequence Core Temperatures are at a Minimum}",
      journal = {Research Notes of the American Astronomical Society},
         year = 2023,
        month = aug,
       volume = {7},
       number = {8},
          eid = {161},
        pages = {161},
          doi = {10.3847/2515-5172/acebd2},
       adsurl = {https://ui.adsabs.harvard.edu/abs/2023RNAAS...7..161F}
}

@ARTICLE{2019ApJ...879..100D,
       author = {{Douglas}, S.~T. and {Curtis}, J.~L. and {Ag{\"u}eros}, M.~A. and {Cargile}, P.~A. and {Brewer}, J.~M. and {Meibom}, S. and {Jansen}, T.},
        title = "{K2 Rotation Periods for Low-mass Hyads and a Quantitative Comparison of the Distribution of Slow Rotators in the Hyades and Praesepe}",
      journal = {\apj},
         year = 2019,
        month = jul,
       volume = {879},
       number = {2},
          eid = {100},
        pages = {100},
          doi = {10.3847/1538-4357/ab2468},
archivePrefix = {arXiv},
       eprint = {1905.06736},
 primaryClass = {astro-ph.SR},
       adsurl = {https://ui.adsabs.harvard.edu/abs/2019ApJ...879..100D}
}

@ARTICLE{2013PASP..125..306F,
       author = {{Foreman-Mackey}, Daniel and {Hogg}, David W. and {Lang}, Dustin and {Goodman}, Jonathan},
        title = "{emcee: The MCMC Hammer}",
      journal = {\pasp},
         year = 2013,
        month = mar,
       volume = {125},
       number = {925},
        pages = {306},
          doi = {10.1086/670067},
archivePrefix = {arXiv},
       eprint = {1202.3665},
 primaryClass = {astro-ph.IM},
       adsurl = {https://ui.adsabs.harvard.edu/abs/2013PASP..125..306F}
}

@ARTICLE{2016ApJ...823...16B,
       author = {{Barnes}, Sydney A. and {Weingrill}, Joerg and {Fritzewski}, Dario and {Strassmeier}, Klaus G. and {Platais}, Imants},
        title = "{Rotation Periods for Cool Stars in the 4 Gyr old Open Cluster M67, The Solar-Stellar Connection, and the Applicability of Gyrochronology to at least Solar Age}",
      journal = {\apj},
         year = 2016,
        month = may,
       volume = {823},
       number = {1},
          eid = {16},
        pages = {16},
          doi = {10.3847/0004-637X/823/1/16},
archivePrefix = {arXiv},
       eprint = {1603.09179},
 primaryClass = {astro-ph.SR},
       adsurl = {https://ui.adsabs.harvard.edu/abs/2016ApJ...823...16B}
}

@ARTICLE{2023A&A...672A.159G,
       author = {{Gruner}, D. and {Barnes}, S.~A. and {Weingrill}, J.},
        title = "{New insights into the rotational evolution of near-solar age stars from the open cluster M 67}",
      journal = {\aap},
         year = 2023,
        month = apr,
       volume = {672},
          eid = {A159},
        pages = {A159},
          doi = {10.1051/0004-6361/202345942},
archivePrefix = {arXiv},
       eprint = {2305.16997},
 primaryClass = {astro-ph.SR},
       adsurl = {https://ui.adsabs.harvard.edu/abs/2023A&A...672A.159G}
}

@ARTICLE{2013AN....334..151W,
       author = {{Wright}, N.~J. and {Drake}, J.~J. and {Mamajek}, E.~E. and {Henry}, G.~W.},
        title = "{The stellar activity-rotation relationship}",
      journal = {Astronomische Nachrichten},
         year = 2013,
        month = feb,
       volume = {334},
       number = {1-2},
        pages = {151},
          doi = {10.1002/asna.201211764},
archivePrefix = {arXiv},
       eprint = {1208.3132},
 primaryClass = {astro-ph.SR},
       adsurl = {https://ui.adsabs.harvard.edu/abs/2013AN....334..151W}
}

@ARTICLE{2022ApJ...933..114D,
       author = {{David}, Trevor J. and {Angus}, Ruth and {Curtis}, Jason L. and {van Saders}, Jennifer L. and {Colman}, Isabel L. and {Contardo}, Gabriella and {Lu}, Yuxi and {Zinn}, Joel C.},
        title = "{Further Evidence of Modified Spin-down in Sun-like Stars: Pileups in the Temperature-Period Distribution}",
      journal = {\apj},
         year = 2022,
        month = jul,
       volume = {933},
       number = {1},
          eid = {114},
        pages = {114},
          doi = {10.3847/1538-4357/ac6dd3},
archivePrefix = {arXiv},
       eprint = {2203.08920},
 primaryClass = {astro-ph.SR},
       adsurl = {https://ui.adsabs.harvard.edu/abs/2022ApJ...933..114D}
}

@ARTICLE{2018ApJ...864..125F,
       author = {{Finley}, Adam J. and {Matt}, Sean P. and {See}, Victor},
        title = "{The Effect of Magnetic Variability on Stellar Angular Momentum Loss. I. The Solar Wind Torque during Sunspot Cycles 23 and 24}",
      journal = {\apj},
         year = 2018,
        month = sep,
       volume = {864},
       number = {2},
          eid = {125},
        pages = {125},
          doi = {10.3847/1538-4357/aad7b6},
archivePrefix = {arXiv},
       eprint = {1808.00063},
 primaryClass = {astro-ph.SR},
       adsurl = {https://ui.adsabs.harvard.edu/abs/2018ApJ...864..125F}
}

@ARTICLE{2017ApJ...850..134S,
       author = {{Somers}, Garrett and {Stauffer}, John and {Rebull}, Luisa and {Cody}, Ann Marie and {Pinsonneault}, Marc},
        title = "{M Dwarf Rotation from the K2 Young Clusters to the Field. I. A Mass-Rotation Correlation at 10 Myr}",
      journal = {\apj},
         year = 2017,
        month = dec,
       volume = {850},
       number = {2},
          eid = {134},
        pages = {134},
          doi = {10.3847/1538-4357/aa93ed},
archivePrefix = {arXiv},
       eprint = {1710.07638},
 primaryClass = {astro-ph.SR},
       adsurl = {https://ui.adsabs.harvard.edu/abs/2017ApJ...850..134S}
}

@ARTICLE{1989ApJ...338..424P,
       author = {{Pinsonneault}, M.~H. and {Kawaler}, Steven D. and {Sofia}, S. and {Demarque}, P.},
        title = "{Evolutionary Models of the Rotating Sun}",
      journal = {\apj},
         year = 1989,
        month = mar,
       volume = {338},
        pages = {424},
          doi = {10.1086/167210},
       adsurl = {https://ui.adsabs.harvard.edu/abs/1989ApJ...338..424P}
}

@ARTICLE{2019ApJ...886..120S,
       author = {{See}, Victor and {Matt}, Sean P. and {Finley}, Adam J. and {Folsom}, Colin P. and {Boro Saikia}, Sudeshna and {Donati}, Jean-Francois and {Fares}, Rim and {H{\'e}brard}, {\'E}lodie M. and {Jardine}, Moira M. and {Jeffers}, Sandra V. and {Marsden}, Stephen C. and {Mengel}, Matthew W. and {Morin}, Julien and {Petit}, Pascal and {Vidotto}, Aline A. and {Waite}, Ian A. and {BCool Collaboration}},
        title = "{Do Non-dipolar Magnetic Fields Contribute to Spin-down Torques?}",
      journal = {\apj},
         year = 2019,
        month = dec,
       volume = {886},
       number = {2},
          eid = {120},
        pages = {120},
          doi = {10.3847/1538-4357/ab46b2},
archivePrefix = {arXiv},
       eprint = {1910.02129},
 primaryClass = {astro-ph.SR},
       adsurl = {https://ui.adsabs.harvard.edu/abs/2019ApJ...886..120S}
}

@ARTICLE{2018MNRAS.474..536S,
       author = {{See}, V. and {Jardine}, M. and {Vidotto}, A.~A. and {Donati}, J.-F. and {Boro Saikia}, S. and {Fares}, R. and {Folsom}, C.~P. and {Jeffers}, S.~V. and {Marsden}, S.~C. and {Morin}, J. and {Petit}, P. and {BCool Collaboration}},
        title = "{The open flux evolution of a solar-mass star on the main sequence}",
      journal = {\mnras},
         year = 2018,
        month = feb,
       volume = {474},
       number = {1},
        pages = {536-546},
          doi = {10.1093/mnras/stx2599},
archivePrefix = {arXiv},
       eprint = {1711.03904},
 primaryClass = {astro-ph.SR},
       adsurl = {https://ui.adsabs.harvard.edu/abs/2018MNRAS.474..536S}
}

@ARTICLE{astropy:2022,
       author = {{Astropy Collaboration} and {Price-Whelan}, Adrian M. and {Lim}, Pey Lian and {Earl}, Nicholas and {Starkman}, Nathaniel and {Bradley}, Larry and {Shupe}, David L. and {Patil}, Aarya A. and {Corrales}, Lia and {Brasseur}, C.~E. and {N{\"o}the}, Maximilian and {Donath}, Axel and {Tollerud}, Erik and {Morris}, Brett M. and {Ginsburg}, Adam and {Vaher}, Eero and {Weaver}, Benjamin A. and {Tocknell}, James and {Jamieson}, William and {van Kerkwijk}, Marten H. and {Robitaille}, Thomas P. and {Merry}, Bruce and {Bachetti}, Matteo and {G{\"u}nther}, H. Moritz and {Aldcroft}, Thomas L. and {Alvarado-Montes}, Jaime A. and {Archibald}, Anne M. and {B{\'o}di}, Attila and {Bapat}, Shreyas and {Barentsen}, Geert and {Baz{\'a}n}, Juanjo and {Biswas}, Manish and {Boquien}, M{\'e}d{\'e}ric and {Burke}, D.~J. and {Cara}, Daria and {Cara}, Mihai and {Conroy}, Kyle E. and {Conseil}, Simon and {Craig}, Matthew W. and {Cross}, Robert M. and {Cruz}, Kelle L. and {D'Eugenio}, Francesco and {Dencheva}, Nadia and {Devillepoix}, Hadrien A.~R. and {Dietrich}, J{\"o}rg P. and {Eigenbrot}, Arthur Davis and {Erben}, Thomas and {Ferreira}, Leonardo and {Foreman-Mackey}, Daniel and {Fox}, Ryan and {Freij}, Nabil and {Garg}, Suyog and {Geda}, Robel and {Glattly}, Lauren and {Gondhalekar}, Yash and {Gordon}, Karl D. and {Grant}, David and {Greenfield}, Perry and {Groener}, Austen M. and {Guest}, Steve and {Gurovich}, Sebastian and {Handberg}, Rasmus and {Hart}, Akeem and {Hatfield-Dodds}, Zac and {Homeier}, Derek and {Hosseinzadeh}, Griffin and {Jenness}, Tim and {Jones}, Craig K. and {Joseph}, Prajwel and {Kalmbach}, J. Bryce and {Karamehmetoglu}, Emir and {Ka{\l}uszy{\'n}ski}, Miko{\l}aj and {Kelley}, Michael S.~P. and {Kern}, Nicholas and {Kerzendorf}, Wolfgang E. and {Koch}, Eric W. and {Kulumani}, Shankar and {Lee}, Antony and {Ly}, Chun and {Ma}, Zhiyuan and {MacBride}, Conor and {Maljaars}, Jakob M. and {Muna}, Demitri and {Murphy}, N.~A. and {Norman}, Henrik and {O'Steen}, Richard and {Oman}, Kyle A. and {Pacifici}, Camilla and {Pascual}, Sergio and {Pascual-Granado}, J. and {Patil}, Rohit R. and {Perren}, Gabriel I. and {Pickering}, Timothy E. and {Rastogi}, Tanuj and {Roulston}, Benjamin R. and {Ryan}, Daniel F. and {Rykoff}, Eli S. and {Sabater}, Jose and {Sakurikar}, Parikshit and {Salgado}, Jes{\'u}s and {Sanghi}, Aniket and {Saunders}, Nicholas and {Savchenko}, Volodymyr and {Schwardt}, Ludwig and {Seifert-Eckert}, Michael and {Shih}, Albert Y. and {Jain}, Anany Shrey and {Shukla}, Gyanendra and {Sick}, Jonathan and {Simpson}, Chris and {Singanamalla}, Sudheesh and {Singer}, Leo P. and {Singhal}, Jaladh and {Sinha}, Manodeep and {Sip{\H{o}}cz}, Brigitta M. and {Spitler}, Lee R. and {Stansby}, David and {Streicher}, Ole and {{\v{S}}umak}, Jani and {Swinbank}, John D. and {Taranu}, Dan S. and {Tewary}, Nikita and {Tremblay}, Grant R. and {de Val-Borro}, Miguel and {Van Kooten}, Samuel J. and {Vasovi{\'c}}, Zlatan and {Verma}, Shresth and {de Miranda Cardoso}, Jos{\'e} Vin{\'\i}cius and {Williams}, Peter K.~G. and {Wilson}, Tom J. and {Winkel}, Benjamin and {Wood-Vasey}, W.~M. and {Xue}, Rui and {Yoachim}, Peter and {Zhang}, Chen and {Zonca}, Andrea and {Astropy Project Contributors}},
        title = "{The Astropy Project: Sustaining and Growing a Community-oriented Open-source Project and the Latest Major Release (v5.0) of the Core Package}",
      journal = {\apj},
         year = 2022,
        month = aug,
       volume = {935},
       number = {2},
          eid = {167},
        pages = {167},
          doi = {10.3847/1538-4357/ac7c74},
archivePrefix = {arXiv},
       eprint = {2206.14220},
 primaryClass = {astro-ph.IM},
       adsurl = {https://ui.adsabs.harvard.edu/abs/2022ApJ...935..167A}
}

@ARTICLE{astropy:2018,
       author = {{Astropy Collaboration} and {Price-Whelan}, A.~M. and {Sip{\H{o}}cz}, B.~M. and {G{\"u}nther}, H.~M. and {Lim}, P.~L. and {Crawford}, S.~M. and {Conseil}, S. and {Shupe}, D.~L. and {Craig}, M.~W. and {Dencheva}, N. and {Ginsburg}, A. and {VanderPlas}, J.~T. and {Bradley}, L.~D. and {P{\'e}rez-Su{\'a}rez}, D. and {de Val-Borro}, M. and {Aldcroft}, T.~L. and {Cruz}, K.~L. and {Robitaille}, T.~P. and {Tollerud}, E.~J. and {Ardelean}, C. and {Babej}, T. and {Bach}, Y.~P. and {Bachetti}, M. and {Bakanov}, A.~V. and {Bamford}, S.~P. and {Barentsen}, G. and {Barmby}, P. and {Baumbach}, A. and {Berry}, K.~L. and {Biscani}, F. and {Boquien}, M. and {Bostroem}, K.~A. and {Bouma}, L.~G. and {Brammer}, G.~B. and {Bray}, E.~M. and {Breytenbach}, H. and {Buddelmeijer}, H. and {Burke}, D.~J. and {Calderone}, G. and {Cano Rodr{\'\i}guez}, J.~L. and {Cara}, M. and {Cardoso}, J.~V.~M. and {Cheedella}, S. and {Copin}, Y. and {Corrales}, L. and {Crichton}, D. and {D'Avella}, D. and {Deil}, C. and {Depagne}, {\'E}. and {Dietrich}, J.~P. and {Donath}, A. and {Droettboom}, M. and {Earl}, N. and {Erben}, T. and {Fabbro}, S. and {Ferreira}, L.~A. and {Finethy}, T. and {Fox}, R.~T. and {Garrison}, L.~H. and {Gibbons}, S.~L.~J. and {Goldstein}, D.~A. and {Gommers}, R. and {Greco}, J.~P. and {Greenfield}, P. and {Groener}, A.~M. and {Grollier}, F. and {Hagen}, A. and {Hirst}, P. and {Homeier}, D. and {Horton}, A.~J. and {Hosseinzadeh}, G. and {Hu}, L. and {Hunkeler}, J.~S. and {Ivezi{\'c}}, {\v{Z}}. and {Jain}, A. and {Jenness}, T. and {Kanarek}, G. and {Kendrew}, S. and {Kern}, N.~S. and {Kerzendorf}, W.~E. and {Khvalko}, A. and {King}, J. and {Kirkby}, D. and {Kulkarni}, A.~M. and {Kumar}, A. and {Lee}, A. and {Lenz}, D. and {Littlefair}, S.~P. and {Ma}, Z. and {Macleod}, D.~M. and {Mastropietro}, M. and {McCully}, C. and {Montagnac}, S. and {Morris}, B.~M. and {Mueller}, M. and {Mumford}, S.~J. and {Muna}, D. and {Murphy}, N.~A. and {Nelson}, S. and {Nguyen}, G.~H. and {Ninan}, J.~P. and {N{\"o}the}, M. and {Ogaz}, S. and {Oh}, S. and {Parejko}, J.~K. and {Parley}, N. and {Pascual}, S. and {Patil}, R. and {Patil}, A.~A. and {Plunkett}, A.~L. and {Prochaska}, J.~X. and {Rastogi}, T. and {Reddy Janga}, V. and {Sabater}, J. and {Sakurikar}, P. and {Seifert}, M. and {Sherbert}, L.~E. and {Sherwood-Taylor}, H. and {Shih}, A.~Y. and {Sick}, J. and {Silbiger}, M.~T. and {Singanamalla}, S. and {Singer}, L.~P. and {Sladen}, P.~H. and {Sooley}, K.~A. and {Sornarajah}, S. and {Streicher}, O. and {Teuben}, P. and {Thomas}, S.~W. and {Tremblay}, G.~R. and {Turner}, J.~E.~H. and {Terr{\'o}n}, V. and {van Kerkwijk}, M.~H. and {de la Vega}, A. and {Watkins}, L.~L. and {Weaver}, B.~A. and {Whitmore}, J.~B. and {Woillez}, J. and {Zabalza}, V. and {Astropy Contributors}},
        title = "{The Astropy Project: Building an Open-science Project and Status of the v2.0 Core Package}",
      journal = {\aj},
         year = 2018,
        month = sep,
       volume = {156},
       number = {3},
          eid = {123},
        pages = {123},
          doi = {10.3847/1538-3881/aabc4f},
archivePrefix = {arXiv},
       eprint = {1801.02634},
 primaryClass = {astro-ph.IM},
       adsurl = {https://ui.adsabs.harvard.edu/abs/2018AJ....156..123A}
}

@ARTICLE{astropy:2013,
       author = {{Astropy Collaboration} and {Robitaille}, Thomas P. and {Tollerud}, Erik J. and {Greenfield}, Perry and {Droettboom}, Michael and {Bray}, Erik and {Aldcroft}, Tom and {Davis}, Matt and {Ginsburg}, Adam and {Price-Whelan}, Adrian M. and {Kerzendorf}, Wolfgang E. and {Conley}, Alexander and {Crighton}, Neil and {Barbary}, Kyle and {Muna}, Demitri and {Ferguson}, Henry and {Grollier}, Fr{\'e}d{\'e}ric and {Parikh}, Madhura M. and {Nair}, Prasanth H. and {Unther}, Hans M. and {Deil}, Christoph and {Woillez}, Julien and {Conseil}, Simon and {Kramer}, Roban and {Turner}, James E.~H. and {Singer}, Leo and {Fox}, Ryan and {Weaver}, Benjamin A. and {Zabalza}, Victor and {Edwards}, Zachary I. and {Azalee Bostroem}, K. and {Burke}, D.~J. and {Casey}, Andrew R. and {Crawford}, Steven M. and {Dencheva}, Nadia and {Ely}, Justin and {Jenness}, Tim and {Labrie}, Kathleen and {Lim}, Pey Lian and {Pierfederici}, Francesco and {Pontzen}, Andrew and {Ptak}, Andy and {Refsdal}, Brian and {Servillat}, Mathieu and {Streicher}, Ole},
        title = "{Astropy: A community Python package for astronomy}",
      journal = {\aap},
         year = 2013,
        month = oct,
       volume = {558},
          eid = {A33},
        pages = {A33},
          doi = {10.1051/0004-6361/201322068},
archivePrefix = {arXiv},
       eprint = {1307.6212},
 primaryClass = {astro-ph.IM},
       adsurl = {https://ui.adsabs.harvard.edu/abs/2013A&A...558A..33A}
}

@ARTICLE{1967ApJ...148..217W,
       author = {{Weber}, Edmund J. and {Davis}, Jr., Leverett},
        title = "{The Angular Momentum of the Solar Wind}",
      journal = {\apj},
         year = 1967,
        month = apr,
       volume = {148},
        pages = {217-227},
          doi = {10.1086/149138},
       adsurl = {https://ui.adsabs.harvard.edu/abs/1967ApJ...148..217W}
}

@ARTICLE{1991ApJ...376..204M,
       author = {{MacGregor}, K.~B. and {Brenner}, M.},
        title = "{Rotational Evolution of Solar-Type Stars. I. Main-Sequence Evolution}",
      journal = {\apj},
         year = 1991,
        month = jul,
       volume = {376},
        pages = {204},
          doi = {10.1086/170269},
       adsurl = {https://ui.adsabs.harvard.edu/abs/1991ApJ...376..204M}
}

@ARTICLE{1997ApJ...480..303K,
       author = {{Krishnamurthi}, Anita and {Pinsonneault}, M.~H. and {Barnes}, S. and {Sofia}, S.},
        title = "{Theoretical Models of the Angular Momentum Evolution of Solar-Type Stars}",
      journal = {\apj},
         year = 1997,
        month = may,
       volume = {480},
       number = {1},
        pages = {303-323},
          doi = {10.1086/303958},
       adsurl = {https://ui.adsabs.harvard.edu/abs/1997ApJ...480..303K}
}

@ARTICLE{1998SSRv...85..161G,
       author = {{Grevesse}, N. and {Sauval}, A.~J.},
        title = "{Standard Solar Composition}",
      journal = {\ssr},
         year = 1998,
        month = may,
       volume = {85},
        pages = {161-174},
          doi = {10.1023/A:1005161325181},
       adsurl = {https://ui.adsabs.harvard.edu/abs/1998SSRv...85..161G}
}

@ARTICLE{2005ApJ...623..585F,
       author = {{Ferguson}, Jason W. and {Alexander}, David R. and {Allard}, France and {Barman}, Travis and {Bodnarik}, Julia G. and {Hauschildt}, Peter H. and {Heffner-Wong}, Amanda and {Tamanai}, Akemi},
        title = "{Low-Temperature Opacities}",
      journal = {\apj},
         year = 2005,
        month = apr,
       volume = {623},
       number = {1},
        pages = {585-596},
          doi = {10.1086/428642},
archivePrefix = {arXiv},
       eprint = {astro-ph/0502045},
 primaryClass = {astro-ph},
       adsurl = {https://ui.adsabs.harvard.edu/abs/2005ApJ...623..585F}
}

@ARTICLE{1996ApJ...456..902R,
       author = {{Rogers}, Forrest J. and {Swenson}, Fritz J. and {Iglesias}, Carlos A.},
        title = "{OPAL Equation-of-State Tables for Astrophysical Applications}",
      journal = {\apj},
         year = 1996,
        month = jan,
       volume = {456},
        pages = {902},
          doi = {10.1086/176705},
       adsurl = {https://ui.adsabs.harvard.edu/abs/1996ApJ...456..902R}
}

@ARTICLE{2002ApJ...576.1064R,
       author = {{Rogers}, F.~J. and {Nayfonov}, A.},
        title = "{Updated and Expanded OPAL Equation-of-State Tables: Implications for Helioseismology}",
      journal = {\apj},
         year = 2002,
        month = sep,
       volume = {576},
       number = {2},
        pages = {1064-1074},
          doi = {10.1086/341894},
       adsurl = {https://ui.adsabs.harvard.edu/abs/2002ApJ...576.1064R}
}

@INPROCEEDINGS{1992IAUS..149..225K,
       author = {{Kurucz}, R.~L.},
        title = "{Model Atmospheres for Population Synthesis}",
    booktitle = {The Stellar Populations of Galaxies},
         year = 1992,
       editor = {{Barbuy}, Beatriz and {Renzini}, Alvio},
       series = {IAU Symposium},
       volume = {149},
        month = jan,
        pages = {225},
       adsurl = {https://ui.adsabs.harvard.edu/abs/1992IAUS..149..225K}
}

@BOOK{1968pss..book.....C,
       author = {{Cox}, J.~P. and {Giuli}, R.~T.},
        title = "{Principles of stellar structure}",
         year = 1968,
       adsurl = {https://ui.adsabs.harvard.edu/abs/1968pss..book.....C}
}

@ARTICLE{2025ApJ...986...59V,
       author = {{Van-Lane}, Phil R. and {Speagle}, Joshua S. and {Eadie}, Gwendolyn M. and {Douglas}, Stephanie T. and {Cargile}, Phillip A. and {Zucker}, Catherine and {Lu}, Yuxi (Lucy) and {Angus}, Ruth},
        title = "{ChronoFlow: A Data-driven Model for Gyrochronology}",
      journal = {\apj},
         year = 2025,
        month = jun,
       volume = {986},
       number = {1},
          eid = {59},
        pages = {59},
          doi = {10.3847/1538-4357/adcd73},
archivePrefix = {arXiv},
       eprint = {2412.12244},
 primaryClass = {astro-ph.SR},
       adsurl = {https://ui.adsabs.harvard.edu/abs/2025ApJ...986...59V}
}

@ARTICLE{2017AJ....154..250L,
       author = {{Lurie}, John C. and {Vyhmeister}, Karl and {Hawley}, Suzanne L. and {Adilia}, Jamel and {Chen}, Andrea and {Davenport}, James R.~A. and {Juri{\'c}}, Mario and {Puig-Holzman}, Michael and {Weisenburger}, Kolby L.},
        title = "{Tidal Synchronization and Differential Rotation of Kepler Eclipsing Binaries}",
      journal = {\aj},
         year = 2017,
        month = dec,
       volume = {154},
       number = {6},
          eid = {250},
        pages = {250},
          doi = {10.3847/1538-3881/aa974d},
archivePrefix = {arXiv},
       eprint = {1710.07339},
 primaryClass = {astro-ph.SR},
       adsurl = {https://ui.adsabs.harvard.edu/abs/2017AJ....154..250L}
}

@ARTICLE{2026arXiv260325792P,
       author = {{Pinsonneault}, Marc H. and {van Saders}, Jennifer L. and {Cao}, Lyra and {Tayar}, Jamie and {Delahaye}, Franck and {Morales}, Leslie M. and {Patton}, Rachel A. and {Rendina}, Matthew C. and {Zinn}, Joel C. and {Claytor}, Zachary R. and {Ash}, Amanda L. and {Byrom}, Susan and {Cao}, Kaili and {Smedile}, Vincent A.},
        title = "{The YREC Stellar Evolution Code: Public Data Release}",
      journal = {arXiv e-prints},
         year = 2026,
        month = mar,
          eid = {arXiv:2603.25792},
        pages = {arXiv:2603.25792},
          doi = {10.48550/arXiv.2603.25792},
archivePrefix = {arXiv},
       eprint = {2603.25792},
 primaryClass = {astro-ph.SR},
       adsurl = {https://ui.adsabs.harvard.edu/abs/2026arXiv260325792P}
}

@ARTICLE{2010Sci...327..977B,
       author = {{Borucki}, William J. and {Koch}, David and {Basri}, Gibor and {Batalha}, Natalie and {Brown}, Timothy and {Caldwell}, Douglas and {Caldwell}, John and {Christensen-Dalsgaard}, J{\o}rgen and {Cochran}, William D. and {DeVore}, Edna and {Dunham}, Edward W. and {Dupree}, Andrea K. and {Gautier}, Thomas N. and {Geary}, John C. and {Gilliland}, Ronald and {Gould}, Alan and {Howell}, Steve B. and {Jenkins}, Jon M. and {Kondo}, Yoji and {Latham}, David W. and {Marcy}, Geoffrey W. and {Meibom}, S{\o}ren and {Kjeldsen}, Hans and {Lissauer}, Jack J. and {Monet}, David G. and {Morrison}, David and {Sasselov}, Dimitar and {Tarter}, Jill and {Boss}, Alan and {Brownlee}, Don and {Owen}, Toby and {Buzasi}, Derek and {Charbonneau}, David and {Doyle}, Laurance and {Fortney}, Jonathan and {Ford}, Eric B. and {Holman}, Matthew J. and {Seager}, Sara and {Steffen}, Jason H. and {Welsh}, William F. and {Rowe}, Jason and {Anderson}, Howard and {Buchhave}, Lars and {Ciardi}, David and {Walkowicz}, Lucianne and {Sherry}, William and {Horch}, Elliott and {Isaacson}, Howard and {Everett}, Mark E. and {Fischer}, Debra and {Torres}, Guillermo and {Johnson}, John Asher and {Endl}, Michael and {MacQueen}, Phillip and {Bryson}, Stephen T. and {Dotson}, Jessie and {Haas}, Michael and {Kolodziejczak}, Jeffrey and {Van Cleve}, Jeffrey and {Chandrasekaran}, Hema and {Twicken}, Joseph D. and {Quintana}, Elisa V. and {Clarke}, Bruce D. and {Allen}, Christopher and {Li}, Jie and {Wu}, Haley and {Tenenbaum}, Peter and {Verner}, Ekaterina and {Bruhweiler}, Frederick and {Barnes}, Jason and {Prsa}, Andrej},
        title = "{Kepler Planet-Detection Mission: Introduction and First Results}",
      journal = {Science},
         year = 2010,
        month = feb,
       volume = {327},
       number = {5968},
        pages = {977},
          doi = {10.1126/science.1185402},
       adsurl = {https://ui.adsabs.harvard.edu/abs/2010Sci...327..977B}
}

@ARTICLE{2019PASP..131a8003M,
       author = {{Masci}, Frank J. and {Laher}, Russ R. and {Rusholme}, Ben and {Shupe}, David L. and {Groom}, Steven and {Surace}, Jason and {Jackson}, Edward and {Monkewitz}, Serge and {Beck}, Ron and {Flynn}, David and {Terek}, Scott and {Landry}, Walter and {Hacopians}, Eugean and {Desai}, Vandana and {Howell}, Justin and {Brooke}, Tim and {Imel}, David and {Wachter}, Stefanie and {Ye}, Quan-Zhi and {Lin}, Hsing-Wen and {Cenko}, S. Bradley and {Cunningham}, Virginia and {Rebbapragada}, Umaa and {Bue}, Brian and {Miller}, Adam A. and {Mahabal}, Ashish and {Bellm}, Eric C. and {Patterson}, Maria T. and {Juri{\'c}}, Mario and {Golkhou}, V. Zach and {Ofek}, Eran O. and {Walters}, Richard and {Graham}, Matthew and {Kasliwal}, Mansi M. and {Dekany}, Richard G. and {Kupfer}, Thomas and {Burdge}, Kevin and {Cannella}, Christopher B. and {Barlow}, Tom and {Van Sistine}, Angela and {Giomi}, Matteo and {Fremling}, Christoffer and {Blagorodnova}, Nadejda and {Levitan}, David and {Riddle}, Reed and {Smith}, Roger M. and {Helou}, George and {Prince}, Thomas A. and {Kulkarni}, Shrinivas R.},
        title = "{The Zwicky Transient Facility: Data Processing, Products, and Archive}",
      journal = {\pasp},
         year = 2019,
        month = jan,
       volume = {131},
       number = {995},
        pages = {018003},
          doi = {10.1088/1538-3873/aae8ac},
archivePrefix = {arXiv},
       eprint = {1902.01872},
 primaryClass = {astro-ph.IM},
       adsurl = {https://ui.adsabs.harvard.edu/abs/2019PASP..131a8003M}
}

@ARTICLE{2012AJ....144..145B,
       author = {{Berta}, Zachory K. and {Irwin}, Jonathan and {Charbonneau}, David and {Burke}, Christopher J. and {Falco}, Emilio E.},
        title = "{Transit Detection in the MEarth Survey of Nearby M Dwarfs: Bridging the Clean-first, Search-later Divide}",
      journal = {\aj},
         year = 2012,
        month = nov,
       volume = {144},
       number = {5},
          eid = {145},
        pages = {145},
          doi = {10.1088/0004-6256/144/5/145},
archivePrefix = {arXiv},
       eprint = {1206.4715},
 primaryClass = {astro-ph.EP},
       adsurl = {https://ui.adsabs.harvard.edu/abs/2012AJ....144..145B}
}

@ARTICLE{2005ApJ...628L.143W,
       author = {{Wood}, B.~E. and {M{\"u}ller}, H.-R. and {Zank}, G.~P. and {Linsky}, J.~L. and {Redfield}, S.},
        title = "{New Mass-Loss Measurements from Astrospheric Ly{\ensuremath{\alpha}} Absorption}",
      journal = {\apjl},
         year = 2005,
        month = aug,
       volume = {628},
       number = {2},
        pages = {L143-L146},
          doi = {10.1086/432716},
archivePrefix = {arXiv},
       eprint = {astro-ph/0506401},
 primaryClass = {astro-ph},
       adsurl = {https://ui.adsabs.harvard.edu/abs/2005ApJ...628L.143W}
}

@ARTICLE{2003A&A...397..147P,
       author = {{Pizzolato}, N. and {Maggio}, A. and {Micela}, G. and {Sciortino}, S. and {Ventura}, P.},
        title = "{The stellar activity-rotation relationship revisited: Dependence of saturated and non-saturated X-ray emission regimes on stellar mass for late-type dwarfs}",
      journal = {\aap},
         year = 2003,
        month = jan,
       volume = {397},
        pages = {147-157},
          doi = {10.1051/0004-6361:20021560},
       adsurl = {https://ui.adsabs.harvard.edu/abs/2003A&A...397..147P}
}

@ARTICLE{McQuillan2014,
       author = {{McQuillan}, A. and {Mazeh}, T. and {Aigrain}, S.},
        title = "{Rotation Periods of 34,030 Kepler Main-sequence Stars: The Full Autocorrelation Sample}",
      journal = {\apjs},
         year = "2014",
        month = "Apr",
       volume = {211},
       number = {2},
          eid = {24},
        pages = {24},
          doi = {10.1088/0067-0049/211/2/24},
archivePrefix = {arXiv},
       eprint = {1402.5694},
 primaryClass = {astro-ph.SR},
       adsurl = {https://ui.adsabs.harvard.edu/abs/2014ApJS..211...24M}
}

@ARTICLE{Santos2019,
       author = {{Santos}, A.~R.~G. and {Garc{\'\i}a}, R.~A. and {Mathur}, S. and
         {Bugnet}, L. and {van Saders}, J.~L. and {Metcalfe}, T.~S. and
         {Simonian}, G.~V.~A. and {Pinsonneault}, M.~H.},
        title = "{Surface Rotation and Photometric Activity for Kepler Targets. I. M and K Main-sequence Stars}",
      journal = {\apjs},
         year = "2019",
        month = "Sep",
       volume = {244},
       number = {1},
          eid = {21},
        pages = {21},
          doi = {10.3847/1538-4365/ab3b56},
archivePrefix = {arXiv},
       eprint = {1908.05222},
 primaryClass = {astro-ph.SR},
       adsurl = {https://ui.adsabs.harvard.edu/abs/2019ApJS..244...21S}
}

@ARTICLE{Irwin2011,
       author = {{Irwin}, Jonathan and {Berta}, Zachory K. and {Burke}, Christopher J. and {Charbonneau}, David and {Nutzman}, Philip and {West}, Andrew A. and {Falco}, Emilio E.},
        title = "{On the Angular Momentum Evolution of Fully Convective Stars: Rotation Periods for Field M-dwarfs from the MEarth Transit Survey}",
      journal = {\apj},
         year = 2011,
        month = jan,
       volume = {727},
       number = {1},
          eid = {56},
        pages = {56},
          doi = {10.1088/0004-637X/727/1/56},
archivePrefix = {arXiv},
       eprint = {1011.4909},
 primaryClass = {astro-ph.SR},
       adsurl = {https://ui.adsabs.harvard.edu/abs/2011ApJ...727...56I}
}

@ARTICLE{Newton2017,
       author = {{Newton}, Elisabeth R. and {Irwin}, Jonathan and {Charbonneau}, David and {Berlind}, Perry and {Calkins}, Michael L. and {Mink}, Jessica},
        title = "{The H{\ensuremath{\alpha}} Emission of Nearby M Dwarfs and its Relation to Stellar Rotation}",
      journal = {\apj},
         year = 2017,
        month = jan,
       volume = {834},
       number = {1},
          eid = {85},
        pages = {85},
          doi = {10.3847/1538-4357/834/1/85},
archivePrefix = {arXiv},
       eprint = {1611.03509},
 primaryClass = {astro-ph.SR},
       adsurl = {https://ui.adsabs.harvard.edu/abs/2017ApJ...834...85N}
}

@ARTICLE{TESS,
       author = {{Ricker}, George R. and {Winn}, Joshua N. and {Vanderspek}, Roland and
         {Latham}, David W. and {Bakos}, G{\'a}sp{\'a}r {\'A}. and
         {Bean}, Jacob L. and {Berta-Thompson}, Zachory K. and
         {Brown}, Timothy M. and {Buchhave}, Lars and {Butler}, Nathaniel R. and
         {Butler}, R. Paul and {Chaplin}, William J. and {Charbonneau}, David and
         {Christensen-Dalsgaard}, J{\o}rgen and {Clampin}, Mark and
         {Deming}, Drake and {Doty}, John and {De Lee}, Nathan and
         {Dressing}, Courtney and {Dunham}, Edward W. and {Endl}, Michael and
         {Fressin}, Francois and {Ge}, Jian and {Henning}, Thomas and
         {Holman}, Matthew J. and {Howard}, Andrew W. and {Ida}, Shigeru and
         {Jenkins}, Jon M. and {Jernigan}, Garrett and {Johnson}, John Asher and
         {Kaltenegger}, Lisa and {Kawai}, Nobuyuki and {Kjeldsen}, Hans and
         {Laughlin}, Gregory and {Levine}, Alan M. and {Lin}, Douglas and
         {Lissauer}, Jack J. and {MacQueen}, Phillip and {Marcy}, Geoffrey and
         {McCullough}, Peter R. and {Morton}, Timothy D. and {Narita}, Norio and
         {Paegert}, Martin and {Palle}, Enric and {Pepe}, Francesco and
         {Pepper}, Joshua and {Quirrenbach}, Andreas and {Rinehart}, Stephen A. and
         {Sasselov}, Dimitar and {Sato}, Bun'ei and {Seager}, Sara and
         {Sozzetti}, Alessandro and {Stassun}, Keivan G. and {Sullivan}, Peter and
         {Szentgyorgyi}, Andrew and {Torres}, Guillermo and {Udry}, Stephane and
         {Villasenor}, Joel},
        title = "{Transiting Exoplanet Survey Satellite (TESS)}",
      journal = {Journal of Astronomical Telescopes, Instruments, and Systems},
         year = "2015",
        month = "Jan",
       volume = {1},
          eid = {014003},
        pages = {014003},
          doi = {10.1117/1.JATIS.1.1.014003},
       adsurl = {https://ui.adsabs.harvard.edu/abs/2015JATIS...1a4003R}
}

@ARTICLE{Green20182,
       author = {{Green}, Gregory M. and {Schlafly}, Edward F. and {Finkbeiner}, Douglas and
         {Rix}, Hans-Walter and {Martin}, Nicolas and {Burgett}, William and
         {Draper}, Peter W. and {Flewelling}, Heather and {Hodapp}, Klaus and
         {Kaiser}, Nicholas and {Kudritzki}, Rolf-Peter and
         {Magnier}, Eugene A. and {Metcalfe}, Nigel and {Tonry}, John L. and
         {Wainscoat}, Richard and {Waters}, Christopher},
        title = "{Galactic reddening in 3D from stellar photometry - an improved map}",
      journal = {\mnras},
         year = 2018,
        month = jul,
       volume = {478},
       number = {1},
        pages = {651-666},
          doi = {10.1093/mnras/sty1008},
archivePrefix = {arXiv},
       eprint = {1801.03555},
 primaryClass = {astro-ph.GA},
       adsurl = {https://ui.adsabs.harvard.edu/abs/2018MNRAS.478..651G}
}

@ARTICLE{Green2018,
       author = {{Green}, {Gregory M.}},
        title = "{dustmaps: A Python interface for maps of interstellar dust}",
      journal = {The Journal of Open Source Software},
         year = "2018",
        month = "Jun",
       volume = {3},
       number = {26},
        pages = {695},
          doi = {10.21105/joss.00695},
       adsurl = {https://ui.adsabs.harvard.edu/abs/2018JOSS....3..695G}
}

@ARTICLE{2003ApJ...586..464B,
       author = {{Barnes}, Sydney A.},
        title = "{On the Rotational Evolution of Solar- and Late-Type Stars, Its Magnetic Origins, and the Possibility of Stellar Gyrochronology}",
      journal = {\apj},
         year = 2003,
        month = mar,
       volume = {586},
       number = {1},
        pages = {464-479},
          doi = {10.1086/367639},
archivePrefix = {arXiv},
       eprint = {astro-ph/0303631},
 primaryClass = {astro-ph},
       adsurl = {https://ui.adsabs.harvard.edu/abs/2003ApJ...586..464B}
}

@ARTICLE{2020AJ....160...90A,
       author = {{Angus}, Ruth and {Beane}, Angus and {Price-Whelan}, Adrian M. and {Newton}, Elisabeth and {Curtis}, Jason L. and {Berger}, Travis and {van Saders}, Jennifer and {Kiman}, Rocio and {Foreman-Mackey}, Daniel and {Lu}, Yuxi Lucy and {Anderson}, Lauren and {Faherty}, Jacqueline K.},
        title = "{Exploring the Evolution of Stellar Rotation Using Galactic Kinematics}",
      journal = {\aj},
         year = 2020,
        month = aug,
       volume = {160},
       number = {2},
          eid = {90},
        pages = {90},
          doi = {10.3847/1538-3881/ab91b2},
archivePrefix = {arXiv},
       eprint = {2005.09387},
 primaryClass = {astro-ph.SR},
       adsurl = {https://ui.adsabs.harvard.edu/abs/2020AJ....160...90A}
}

@ARTICLE{1984ApJ...287..769N,
       author = {{Noyes}, R.~W. and {Weiss}, N.~O. and {Vaughan}, A.~H.},
        title = "{The relation between stellar rotation rate and activity cycle periods.}",
      journal = {\apj},
         year = 1984,
        month = dec,
       volume = {287},
        pages = {769-773},
          doi = {10.1086/162735},
       adsurl = {https://ui.adsabs.harvard.edu/abs/1984ApJ...287..769N}
}

@ARTICLE{2010A&A...510A..72F,
       author = {{Fedele}, D. and {van den Ancker}, M.~E. and {Henning}, Th. and {Jayawardhana}, R. and {Oliveira}, J.~M.},
        title = "{Timescale of mass accretion in pre-main-sequence stars}",
      journal = {\aap},
         year = 2010,
        month = feb,
       volume = {510},
          eid = {A72},
        pages = {A72},
          doi = {10.1051/0004-6361/200912810},
archivePrefix = {arXiv},
       eprint = {0911.3320},
 primaryClass = {astro-ph.SR},
       adsurl = {https://ui.adsabs.harvard.edu/abs/2010A&A...510A..72F}
}

@ARTICLE{2019ApJ...885L..30F,
       author = {{Finley}, Adam J. and {Hewitt}, Amy L. and {Matt}, Sean P. and {Owens}, Mathew and {Pinto}, Rui F. and {R{\'e}ville}, Victor},
        title = "{Direct Detection of Solar Angular Momentum Loss with the Wind Spacecraft}",
      journal = {\apjl},
         year = 2019,
        month = nov,
       volume = {885},
       number = {2},
          eid = {L30},
        pages = {L30},
          doi = {10.3847/2041-8213/ab4ff4},
archivePrefix = {arXiv},
       eprint = {1910.10177},
 primaryClass = {astro-ph.SR},
       adsurl = {https://ui.adsabs.harvard.edu/abs/2019ApJ...885L..30F}
}

@ARTICLE{2015A&A...577A..98G,
       author = {{Gallet}, F. and {Bouvier}, J.},
        title = "{Improved angular momentum evolution model for solar-like stars. II. Exploring the mass dependence}",
      journal = {\aap},
         year = 2015,
        month = may,
       volume = {577},
          eid = {A98},
        pages = {A98},
          doi = {10.1051/0004-6361/201525660},
archivePrefix = {arXiv},
       eprint = {1502.05801},
 primaryClass = {astro-ph.SR},
       adsurl = {https://ui.adsabs.harvard.edu/abs/2015A&A...577A..98G}
}

@ARTICLE{1984ApJ...279..763N,
       author = {{Noyes}, R.~W. and {Hartmann}, L.~W. and {Baliunas}, S.~L. and {Duncan}, D.~K. and {Vaughan}, A.~H.},
        title = "{Rotation, convection, and magnetic activity in lower main-sequence stars.}",
      journal = {\apj},
         year = 1984,
        month = apr,
       volume = {279},
        pages = {763-777},
          doi = {10.1086/161945},
       adsurl = {https://ui.adsabs.harvard.edu/abs/1984ApJ...279..763N}
}

@ARTICLE{2005MNRAS.360..458B,
       author = {{Badnell}, N.~R. and {Bautista}, M.~A. and {Butler}, K. and {Delahaye}, F. and {Mendoza}, C. and {Palmeri}, P. and {Zeippen}, C.~J. and {Seaton}, M.~J.},
        title = "{Updated opacities from the Opacity Project}",
      journal = {\mnras},
         year = 2005,
        month = jun,
       volume = {360},
       number = {2},
        pages = {458-464},
          doi = {10.1111/j.1365-2966.2005.08991.x},
archivePrefix = {arXiv},
       eprint = {astro-ph/0410744},
 primaryClass = {astro-ph},
       adsurl = {https://ui.adsabs.harvard.edu/abs/2005MNRAS.360..458B}
}

@ARTICLE{2023ApJ...947L...3B,
       author = {{Bouma}, Luke G. and {Palumbo}, Elsa K. and {Hillenbrand}, Lynne A.},
        title = "{The Empirical Limits of Gyrochronology}",
      journal = {\apjl},
         year = 2023,
        month = apr,
       volume = {947},
       number = {1},
          eid = {L3},
        pages = {L3},
          doi = {10.3847/2041-8213/acc589},
archivePrefix = {arXiv},
       eprint = {2303.08830},
 primaryClass = {astro-ph.SR},
       adsurl = {https://ui.adsabs.harvard.edu/abs/2023ApJ...947L...3B}
}

@ARTICLE{2026A&A...706A.262S,
       author = {{Spada}, F. and {Lanzafame}, A.~C.},
        title = "{Rotational evolution of slow─rotator sequence stars: II. Modeling the wind braking and rotational coupling in the entire mass range of solar-like stars}",
      journal = {\aap},
         year = 2026,
        month = feb,
       volume = {706},
          eid = {A262},
        pages = {A262},
          doi = {10.1051/0004-6361/202557731},
archivePrefix = {arXiv},
       eprint = {2512.12782},
 primaryClass = {astro-ph.SR},
       adsurl = {https://ui.adsabs.harvard.edu/abs/2026A&A...706A.262S}
}

@ARTICLE{2020A&A...636A..76S,
       author = {{Spada}, F. and {Lanzafame}, A.~C.},
        title = "{Competing effect of wind braking and interior coupling in the rotational evolution of solar-like stars}",
      journal = {\aap},
         year = 2020,
        month = apr,
       volume = {636},
          eid = {A76},
        pages = {A76},
          doi = {10.1051/0004-6361/201936384},
archivePrefix = {arXiv},
       eprint = {1908.00345},
 primaryClass = {astro-ph.SR},
       adsurl = {https://ui.adsabs.harvard.edu/abs/2020A&A...636A..76S}
}

@ARTICLE{2017ApJ...845...46F,
       author = {{Finley}, Adam J. and {Matt}, Sean P.},
        title = "{The Effect of Combined Magnetic Geometries on Thermally Driven Winds. I. Interaction of Dipolar and Quadrupolar Fields}",
      journal = {\apj},
         year = 2017,
        month = aug,
       volume = {845},
       number = {1},
          eid = {46},
        pages = {46},
          doi = {10.3847/1538-4357/aa7fb9},
archivePrefix = {arXiv},
       eprint = {1707.04078},
 primaryClass = {astro-ph.SR},
       adsurl = {https://ui.adsabs.harvard.edu/abs/2017ApJ...845...46F}
}

@ARTICLE{2018ApJ...854...78F,
       author = {{Finley}, Adam J. and {Matt}, Sean P.},
        title = "{The Effect of Combined Magnetic Geometries on Thermally Driven Winds. II. Dipolar, Quadrupolar, and Octupolar Topologies}",
      journal = {\apj},
         year = 2018,
        month = feb,
       volume = {854},
       number = {2},
          eid = {78},
        pages = {78},
          doi = {10.3847/1538-4357/aaaab5},
archivePrefix = {arXiv},
       eprint = {1801.07662},
 primaryClass = {astro-ph.SR},
       adsurl = {https://ui.adsabs.harvard.edu/abs/2018ApJ...854...78F}
}
\bibliographystyle{aasjournal}

\end{document}